\documentclass[11pt]{article}

\usepackage[hidelinks]{hyperref}
\hypersetup{colorlinks=true, allcolors=[rgb]{0,0,1}}
\usepackage{graphicx}
\usepackage{listings}
\usepackage{amsmath}
\usepackage{amsthm}
\usepackage{amssymb}
\usepackage{algorithm}
\usepackage{algpseudocode}
\usepackage{setspace}
\usepackage{comment}
\usepackage[dvipsnames]{xcolor}
\usepackage{natbib}
\newcommand\numberthis{\addtocounter{equation}{1}\tag{\theequation}}

\theoremstyle{definition}

\newtheorem{runnexample}{Running Example}[section]
\newcommand{\D}{\displaystyle}

\newcommand\mc{\mathcal}
\newcommand\mbb{\mathbb}

\providecommand{\keywords}[1]
{
  \small	
  \textbf{\textit{Keywords---}} #1
}

\usepackage{geometry}

\title{Hamiltonian Monte Carlo for (Physics) Dummies}
\author{Arghya Mukherjee\\ Department of Mathematics and Statistics \\ IIT Kanpur \\ \texttt{arghyam21@iitk.ac.in} \and Dootika Vats \\ Department of Mathematics and Statistics \\ IIT Kanpur \\ \texttt{dootika@iitk.ac.in}}

\date{}

\begin{document}
\maketitle
\begin{abstract}
Sampling-based inference has seen a surge of interest in recent years. Hamiltonian Monte Carlo (HMC) has emerged as a powerful algorithm that leverages concepts from Hamiltonian dynamics to efficiently explore complex target distributions. Variants of HMC are available in popular software packages, enabling off-the-shelf implementations that have greatly benefited the statistics and machine learning communities. At the same time, the availability of such black-box implementations has made it challenging for users to understand the inner workings of HMC, especially when they are unfamiliar with the underlying physical principles. We provide a pedagogical overview of HMC that aims to bridge the gap between its theoretical foundations and practical applicability. This review article seeks to make HMC more accessible to applied researchers by highlighting its advantages, limitations, and role in enabling scalable and exact Bayesian inference for complex models.
\end{abstract}



\section{Introduction}

Markov chain Monte Carlo (MCMC) methods lie at the heart of Bayesian computation, providing a general mechanism to sample from complex posterior distributions that are analytically intractable. Yet, standard algorithms such as the Metropolis–Hastings \citep{hastings1970monte,metropolis1953equation} or the Gibbs sampler \citep{geman1984stochastic} often struggle in high dimensions or when model parameters exhibit strong correlations, leading to slow mixing and poor scalability in modern applications. Hamiltonian Monte Carlo (HMC),  introduced by \cite{duane1987hybrid} and made popular by \cite{neal2011mcmc}, has been a breakthrough algorithm providing a sampling solution that overcomes many of these hurdles. A standard HMC algorithm introduces auxiliary variables and employs Hamiltonian dynamics to obtain a next candidate draw.  Despite its elegance, the practical use of HMC entails many challenges. Effective deployment requires careful tuning of several hyperparameters that govern both numerical stability and sampling efficiency. Recent investment in off-the-shelf adapted implementations of HMC, like Stan \citep{stan:dev:team:2017} and PyMC3 \citep{salvatier2016probabilistic}, have made it significantly easier for practitioners to employ adaptive HMC algorithms. On the one hand, this has led to renewed interest in Bayesian modeling strategies across disciplines, while on the other hand, HMC and its variants are often treated as black-box algorithms, limiting the opportunities for practitioners to understand and utilize them effectively.

Recently, a few efforts have been made to exposit the details of an HMC algorithm. \cite{neal2011mcmc}  {\color{black}presents} a great first introduction, \cite{betancourt:2017} {\color{black}provides} elegant explanations {\color{black}providing} detailed heuristics and pedagogical motivations. \cite{granados:2025} focus on explaining HMC using physics fundamentals, and \cite{thomas:2021} walk a reader through critical implementation details in \texttt{R}. However, arguably, all works assume a certain amount of knowledge of physics. Despite these excellent works, the authors of this paper still found it challenging to intuitively understand the fine interplay between Hamiltonian dynamics and Monte Carlo sampling. Further, it remained unclear how HMC fits into a traditional Metropolis-Hastings setup. This work presents a pedagogical explanation of HMC, introducing Hamiltonian dynamics in a way that, hopefully, does not require much beyond elementary physics. The intended audience is researchers in statistics and machine learning who are interested in sampling methods, understand the fundamentals of Metropolis-Hastings, but, like the authors of this paper, are ``physics dummies".

The rest of the paper is organized as follows. In Section~\ref{sec:MCMC}, we present the Metropolis-Hastings algorithm and its Jacobian-based extension, the Metropolis-Hastings-Green method, from which we lay the foundation of HMC in Section~\ref{sec:HMC}. An effective HMC implementation requires careful choice of tuning parameters, heuristics and theory for which are outlined in Section~\ref{sec:optimal_tuning}. Section~\ref{sec:data_analysis} demonstrates the implementation of HMC {\color{black} for a Bayesian model applied to a} real dataset. Our approach in these sections is to keep the discussion fairly pedagogical to ensure clarity. In Section~\ref{sec:HMC_variants} we assemble and present recent developments in HMC and its variants, and we end with a discussion in Section~\ref{sec:discussion}.

\section{Markov chain Monte Carlo}
\label{sec:MCMC}

A typical MCMC problem may be posed in the following way. Interest is in obtaining samples from a distribution with density $\pi(x)$, defined on a state space $\mathcal{X} \subseteq \mathbb{R}^d$. (We employ the common abuse of notation, referring to $\pi$ as both the density and the distribution.)  In many Bayesian applications, $\pi$ represents the posterior density of the model parameters given the observed data. However, the functional form of $\pi$ often does not correspond to any known distribution, and it is typically available only up to a normalizing constant. That is, 
\[
  \pi(x) \propto \tilde{\pi}(x)\,,
\]
such that $\tilde{\pi}(x)$ is completely known and the normalizing constant is possibly unknown. Consequently, obtaining independent samples from $\pi$ can be difficult even for moderately large $d$. MCMC provides a general framework for sampling from such intractable target densities by constructing a Markov chain whose stationary and limiting distribution is $\pi$. An appropriately constructed MCMC algorithm would ideally be \textit{ergodic}, in that it is guaranteed to converge to $\pi$ in total variation distance, irrespective of the starting value of the Markov chain. Although the construction of such algorithms might seem daunting, due to the path-breaking work of \cite{metropolis1953equation}, this is not so challenging. The challenge is to construct algorithms that exhibit this convergence rapidly.

\subsection{Metropolis-Hastings algorithm}\label{sec:MH}

The Metropolis algorithm originated from a physics application in the 1950s \citep{metropolis1953equation}, and was further extended nearly two
decades later by \cite{hastings1970monte}, thus giving rise to the Metropolis–Hastings (MH) algorithm\footnote{More recently, it has {\color{black}been highlighted} that much of the laborious programming implementation in the original paper was carried out by Arianna Rosenbluth, and the name should possibly be changed to the Metropolis-Rosenbluth-Teller-Hastings Algorithm. Although we keep the common name in this paper, we direct the reader to \cite{rosenbluth2022}, a wonderful article on the contributions of Arianna Rosenbluth.}. The MH algorithm constructs a Markov chain using a user-chosen proposal distribution that generates a candidate sample based on the current state. This candidate is then accepted or rejected based on a certain probability that guarantees $\pi$ as the stationary distribution. The specifics are provided in Algorithm~\ref{alg:mh}, which employs a proposal distribution $Q$ with conditional density $q(\cdot \mid x)$ to generate a candidate sample $x^*$. The acceptance function
\begin{equation}
  \label{eq:mh_acceptance}
    \alpha_{\text{MH}}(x^{(t)}, x^*) = \min \left\{1,\dfrac{\pi({x}^*)}{\pi({x}^{(t)})} \dfrac{q({x}^{(t)} \,|\, {x}^*)}{q({x}^* \,|\, {x}^{(t)})} \right\}\,,
\end{equation}
is the MH acceptance function that keeps $\pi$ invariant for this algorithm.
\begin{algorithm}
\caption{Metropolis--Hastings algorithm}
\label{alg:mh}
\begin{algorithmic}[1]
\State \textbf{Initialize:} Set initial value ${x}^{(0)}$
\For{$t = 0, 1, 2, \ldots, T-1$}
    \State Draw a proposal ${x}^* \sim q(\,\cdot\,|\,{x}^{(t)})$, where $q$ is a proposal density.
    \State Draw $W \sim \text{Uniform}(0, 1)$ independently.
    \State Compute the acceptance ratio:
    \[
        r({x}^{(t)}, {x}^*) 
        = 
        \dfrac{\pi({x}^*)}{\pi({x}^{(t)})}
        \dfrac{q({x}^{(t)} \,|\, {x}^*)}
              {q({x}^* \,|\, {x}^{(t)})}.
    \]
    \If{$W \leq \min\{1, r({x}^{(t)}, {x}^*)\}$}
        \State Set ${x}^{(t+1)} = {x}^*$
    \Else
        \State Set ${x}^{(t+1)} = {x}^{(t)}$
    \EndIf
\EndFor
\State \textbf{Return:} $\{{x}^{(t)}\}_{t=0}^{T-1}$.
\end{algorithmic}
\end{algorithm}

The effectiveness of an MH algorithm relies almost entirely on the choice of the proposal distribution, $Q$. A ``good'' proposal should often propose values in high probability areas of $\pi$ and should also sufficiently explore the tails. Consequently, a ``good'' proposal makes large moves in the state-space, $\mathcal{X}$. A ``bad'' proposal is one that proposes values close to the current value, yielding low movement of the sampler. 

Significant effort has gone into constructing effective sampling strategies. Perhaps the easiest one to implement is the original random walk Metropolis (RWM) of \cite{metropolis1953equation} where $Q \equiv N(x^{(t)}, h)$ for a tunable $h > 0$. RWM algorithms are often a practitioner's go-to choice, requiring very little overhead and guaranteeing quick implementation time. However, the proposal struggles to move rapidly in the state-space, particularly in high dimensions, as it utilizes no information from $\pi$ itself on where high-probability areas might be. This led to the development of powerful gradient-based proposals like the Metropolis Adjusted Langevin Algorithm (MALA) \citep{roberts1996exponential} 
\begin{equation}
  \label{eq:MALA}
Q \equiv N \left(x^{(t)} + \dfrac{h}{2} \nabla \log \pi(x), h \mathbb{I}_d\right)\,,
\end{equation}
where $h > 0$ is tunable. The MALA proposal moves intentionally towards areas of high probability, learning from the local geometry. The result is an improved exploration of the state-space in high dimensions. HMC is also a gradient-based algorithm that cleverly uses multiple gradients to propose the next move of the sampler. However, HMC does not very elegantly fit into a typical MH scheme and can be viewed as an instance of the general Metropolis-Hastings-Green algorithm.

\subsection{Metropolis-Hastings-Green algorithm}\label{sec:MHG}

Consider a situation when we might want to employ a proposal that is \textit{deterministic}. That is, there is nothing random about the proposal. Certainly, this will be extremely ineffective as the purpose of MCMC sampling is to obtain \textit{random} draws from $\pi$. \cite{green:1995}, however, identified that one may augment the state-space in such a way that a deterministic proposal in an augmented state-space is in fact random in the original state-space. This is done through the Metropolis–Hastings–Green algorithm with Jacobians (MHGJ)\footnote{The Metropolis-Hastings-Green algorithm has found incredible use in transdimensional MCMC algorithms, and its connections to HMC are somewhat less known. Here, we present only what is essential for their use in HMC.}.

Consider a conditional distribution $S(\cdot \mid x)$ with density $s(\cdot \mid x)$ defined on $\mathcal{Y}$ such that together with $\pi$, this yields the joint density
\begin{equation}
  \label{eq:mhgreen_joint}
  \pi_{\text{Aug}}(x,y) = \pi(x) s(y \mid x)\,.
\end{equation}
Since $s$ is a valid density, integrating out $y$ from $\pi_{\text{Aug}}$ will yield our $\pi$ of interest. In an MHGJ algorithm, the sampler will attempt to keep $\pi_{\text{Aug}}$ invariant, however, we will only be retaining the $x$ samples (in effect, integrating out $y$). The state-space for the augmented distribution is $\mathcal{Z} = \mathcal{X} \times \mathcal{Y}$. Typically, $S(\cdot \mid x)$ is chosen in such a way that sampling directly from it is straightforward. 

The key idea in MHGJ is then the following: given a state $(x^{(t)}, y^{(t)})$, we first draw a fresh $y$ from the true conditional distribution $S(\cdot \mid x^{(t)})$, and then apply a \textit{deterministic} mapping on the \textit{random} $(x^{(t)}, y)$, yielding $(x^*, y^*)$. This new state, $(x^*, y^*)$, is then accepted or rejected based on some acceptance probability that is designed to keep $\pi_{\text{Aug}}$ invariant. Since we ignore the $y$ samples, the marginal sampler will be $\pi$-invariant. The deterministic mapping here is a central piece and must satisfy certain properties. We denote this function by $g$ and require that $g: \mc{Z} \to \mc{Z}$ satisfies $g(g(z)) = z$ for all $z \in \mathcal{Z}$\footnote{Such a function is called an involution.} and the Jacobian of $g$ is everywhere non-singular. We present the MHGJ algorithm in Algorithm~\ref{alg:mhg}.

\begin{algorithm}
\caption{Metropolis--Hastings--Green with Jacobians (MHGJ) Algorithm}
\label{alg:mhg}
\begin{algorithmic}[1]
\State \textbf{Initialize:} Set initial value ${x}^{(0)}$
\For{$t = 0, 1, 2, \ldots, T-1$}
    \State Draw the auxiliary variable ${y} \sim S(\,\cdot\,|\,{x}^{(t)})$ with density $s({y} \,|\, {x}^{(t)})$.
    \State Apply the involution $g$, and obtain the proposed state
   \[
   ({x}^*, {y}^*) = g({x}^{(t)}, {y})\,.
  \]
    \State Compute the acceptance ratio:
   $$ r({x}^{(t)}, {x}^*) 
        = 
        \dfrac{\pi({x}^*)\, s({y}^* \,|\, {x}^*)}
               {\pi({x}^{(t)})\, s({y} \,|\, {x}^{(t)})}
        \, \big| \det\big(\nabla g({x}^{(t)}, {y})\big) \big|. $$
    \State Draw $W \sim \text{Uniform}(0, 1)$ independently.
    \If{$W \leq \min\{1, r({x}^{(t)}, {x}^*)\}$}
        \State Set ${x}^{(t+1)} = {x}^*$
    \Else
        \State Set ${x}^{(t+1)} = {x}^{(t)}$
    \EndIf
\EndFor
\State \textbf{Return:} $\{{x}^{(t)}\}_{t=0}^{T-1}$ as the Markov chain samples.
\end{algorithmic}
\end{algorithm}

Algorithm~\ref{alg:mhg} {\color{black}seems to employ}  an acceptance function noticeably different from standard MH.  This is not entirely true, as we will now see. First, we note that Algorithm~\ref{alg:mhg} should be viewed as a composition of two Markov kernels.  The first kernel does a Gibbs step in Step~3, which keeps $\pi_{\text{Aug}}$ invariant like any Gibbs full conditional update. The second kernel is comprised of Steps 4-11 for which the input is $(x^{(t)}, y)$. Now, given this input, Step~4 acts as a proposal step where the proposal is indeed \textit{deterministic}. For Step~5, see that,
\begin{align*}
r({x}^{(t)}, {x}^*) & = \dfrac{\pi({x}^*)\, s({y}^* \,|\, {x}^*)}{\pi({x}^{(t)})\, s({y} \,|\, {x}^{(t)})}\, \big| \det\big(\nabla g({x}^{(t)}, {y})\big) \big| \\
& = \dfrac{\pi_{\text{Aug}}(x^*, y^*) }{\pi_{\text{Aug}}(x^{(t)}, y) } \,\big| \det\big(\nabla g({x}^{(t)}, {y})\big) \big|\,. \numberthis \label{eq:mhg_acceptance}
\end{align*}
This restructuring makes the acceptance function somewhat similar to \eqref{eq:mh_acceptance}, where the ratio of the target density at the proposed and the current points also appears. Since the proposal is a Dirac mass, only the Jacobian of the transformation remains in place of the ratio of the proposal densities in \eqref{eq:mh_acceptance}. See \cite{tier:1998} for more details on this last comment.

For the algorithm to be correct, $g$ must be involutive, acting as its own inverse, ensuring reversibility of the deterministic proposal mechanism within the MHGJ framework. \cite{glatt2026sacred,neklyudov:2020} present a holistic ``involutive MCMC" paradigm, which shows that a wide class of MCMC kernels can be expressed via a self‐inverse map $g$ acting on an augmented state-space. Subsequent work has extended this idea to non-parametric models \citep{mak:2022}, adaptive involutive map \citep{liu:2024}, and to learned volume-preserving neural involutions \citep{spanbauer:2020}. 

Implementing the MHGJ algorithm thus entails specifying a suitable auxiliary distribution $S$ and a deterministic map $g$. HMC does exactly this by choosing $S$ and $g$ by drawing connections with  Hamiltonian dynamics. For most of Section~\ref{sec:HMC}, we assume the dimension $d = 1$ for ease of explanation. A general algorithm for a $d$-dimensional target is discussed in Section~\ref{sub:hmc_higher_dim}.

\section{Hamiltonian Monte Carlo}
\label{sec:HMC}

To address computational challenges in lattice quantum chromodynamics, \cite{duane1987hybrid} stitched molecular dynamics with MCMC, introducing the ``Hybrid Monte Carlo'' method. Since its introduction to statistics by \cite{neal1996priors} in neural network modeling, ``Hybrid Monte Carlo'' has become very popular and was later adopted as ``Hamiltonian Monte Carlo'' as a generic tool for sampling smooth distributions in Bayesian statistical inference.
The Hamiltonian approach offers an alternative methodology for describing motion in particle systems and has served as the foundation for the development of HMC in the statistics community. {\color{black}We begin by first describing motion via Hamiltonian dynamics.}

\subsection{Hamiltonian dynamics}
\label{sec:Hamiltonian Dynamics}

Imagine a ball in a large, frictionless one-dimensional bowl or valley. We may place the ball at any point in the bowl and give it a “push” to initiate its motion. Because the bowl is frictionless, the ball continues to move down and up the bowl according to its initial momentum. After descending into the bowl, it travels up the opposite side, gradually slowing as it climbs. At a sufficiently high point, the ball momentarily comes to rest before sliding back down the side of the bowl. The act of sliding back down corresponds to a change in the sign of the momentum. If the initial ``push'' is smaller, the highest point the ball can reach in the bowl is correspondingly lower. This behavior arises because the ball moves while conserving its total energy, which is determined jointly by the initial position and the momentum. As the ball moves up the sides of the bowl, its kinetic energy decreases (as its momentum reduces) while its potential energy increases (as its height increases), maintaining a constant total energy. A larger initial push gives the ball greater kinetic energy, allowing it to travel farther up the sides of the bowl.

The reader may further imagine that this system is simple enough that if a physicist were to close their eyes at some point while observing the ball, they would still be able to accurately predict the location of the ball after some time 
$s$. This is because the motion of the ball in the bowl follows well-defined laws that can be described using Hamiltonian dynamics. As we proceed in this section, we invite the reader to keep this ball-and-bowl analogy in mind to help build intuition.

Consider, now, the motion of an object on any frictionless surface. At any time $t$, let $x_t$ denote the position of the particle and let $p_t$ denote its momentum. At any given moment, an object possesses both potential energy and kinetic energy. The potential energy, $U(x_t)$, is determined by the position, and the kinetic energy, $K(p_t)$, is determined by its momentum.  In the absence of external forces or friction, the object evolves in such a way that its total energy remains conserved. That is,
\[
H(x_t,p_t) := U(x_t) + K(p_t)
\]
remains constant as the object moves and is called the Hamiltonian. The actual value of the Hamiltonian is then dictated by the initial condition $(x_0, p_0)$, where the object started from and with what momentum. The conservation of the Hamiltonian is incredibly useful in determining how the object will move in a given system. Given  $U$ and $K$ for a system, the temporal evolution of the object is governed by Hamilton's equations:
\begin{equation}\label{eq:hamiltonian}
    \dfrac{dx_t}{dt} = \dfrac{\partial H(x_t,p_t)}{\partial p_t} \qquad \text{ and } \qquad \dfrac{dp_t}{dt} = - \dfrac{\partial H(x_t,p_t)}{\partial x_t}\,.
\end{equation}
That is, \eqref{eq:hamiltonian} informs the change in the object's position and momentum with time. Given an $(x_0, p_0)$, if an object moves according to Hamilton's equations, then the Hamiltonian is conserved and $H(x_0, p_0) = H(x_t, p_t)$ for all $t$. Further, the solution to the partial differential equations in \eqref{eq:hamiltonian} allows for a functional relationship of the state of the object at any time $t$.

\begin{runnexample}
\label{ex:toy_example}

 Consider a system where the potential, kinetic, and total (Hamiltonian) energies are expressed as  
\[
U(x) = \dfrac{x^2}{2}, 
\qquad 
K(p) = \dfrac{p^2}{2}, 
\qquad 
H(x,p) = \dfrac{x^2}{2} + \dfrac{p^2}{2}\,.
\]
Applying Hamilton’s equations, we obtain the system of differential equations  
\begin{equation}\label{eq:gaussian_hamiltonian}
    \dfrac{dx_t}{dt} = \dfrac{\partial H(x_t,p_t)}{\partial p_t} = p_t,
\qquad 
\dfrac{dp_t}{dt} = -\,\dfrac{\partial H(x_t,p_t)}{\partial x_t} = -x_t\,.
\end{equation}
For initial condition $(x_0, p_0)$, the system of partial differential equations in \eqref{eq:gaussian_hamiltonian} admits the analytical solution
\begin{equation}\label{eq:gaussian_motion_eq}
    x_t = x_0 \cos t + p_0 \sin t \qquad  \text{ and }
\qquad 
p_t = -x_0 \sin t + p_0 \cos t\,.
\end{equation}
An analytical solution, as in \eqref{eq:gaussian_motion_eq}, allows the knowledge of the state of the object at any given time point $t$. This is eventually what will be used in the HMC sampler.
\end{runnexample}

\subsection{Ideal HMC}\label{sec:ideal_HMC}

\begin{figure}
    \centering
 \includegraphics[width=\linewidth]{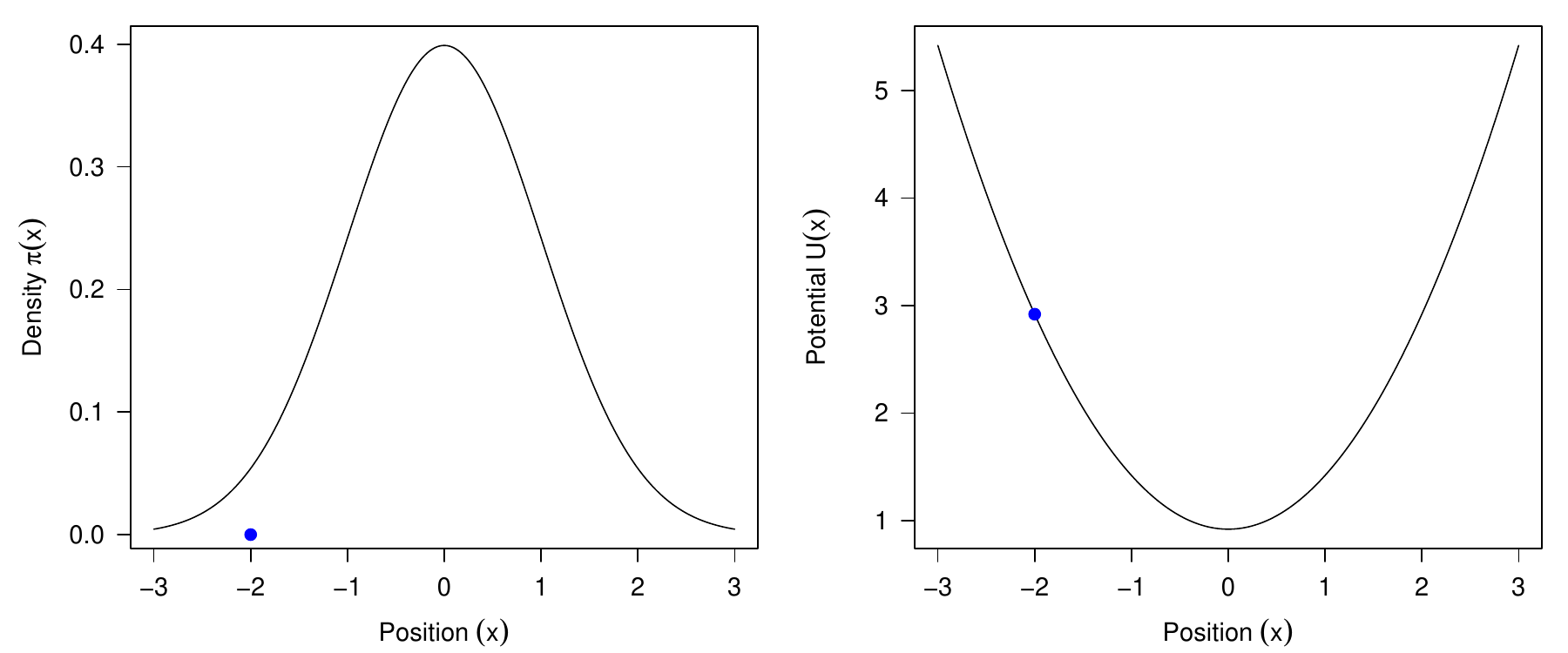}
    \caption{An equivalent representation of Gaussian potential energy.}
    \label{fig:gaussian_pe}
\end{figure}

A key idea in HMC is to assume that the target density corresponds to a potential energy function. Specifically, we set
\[
U(x) = - \log\pi(x)\,.
\]
Then a draw $X = x \sim \pi$ can be interpreted as a state possessing potential energy $U(x)$. Points located in the tails of the distribution correspond to regions of high potential energy, and points in the center of the distribution correspond to having low potential energy; see Figure~\ref{fig:gaussian_pe}. This might seem counterintuitive, but note that due to conservation of energy, high potential energy in the tails implies low kinetic energy. At a far enough point in the tail, the kinetic energy will be zero, at which point the direction of momentum will flip, making the object travel back to areas of high probability. 
If the object is assumed to have mass $m$, the kinetic energy function is
 \[
 K(p) = \dfrac{p^2}{2m}\,.
 \]
 Since we are setting up an imaginary system (there is no real object in motion), we may assume any form of kinetic energy; however, the above choice is standard.

\begin{runnexample}[Normal distribution]
    Suppose $\pi \equiv N(0,1)$, then $U(x) = x^2/2$ and Figure~\ref{fig:gaussian_pe} plots this potential energy function. We may assume an associated kinetic energy of $K(p) = p^2/2$ so that $m  = 1$. This yields the same Hamiltonian as in Example~\ref{ex:toy_example}:
    \[
    H(x,p) = \dfrac{x^2}{2} + \dfrac{p^2}{2}\,.
    \]
Given an initial condition $(x_0,p_0)$, the object will evolve according to the Hamiltonian dynamics \eqref{eq:gaussian_hamiltonian} conserving energy in its path. Since an analytical solution is available in \eqref{eq:gaussian_motion_eq}, we can draw the path of the object as time evolves, as a function of $x$ and $p$. In Figure~\ref{fig:time_trajectory}, the trajectory illustrates how the system evolves when initialized at $x_0 = 2$ and $p_0 = -2$. The particle begins its motion from this state and moves along a circular trajectory in the $(x,p)$ plane, conserving its total energy throughout. Because the total energy is preserved for a given initial condition, different starting values of $(x_0, p_0)$ correspond to distinct energy contours. In Figure \ref{fig:time_traj_diff_momentum}, we consider a second trajectory with the same starting position, $x_0$, but a different initial momentum, $p_0 = -1$. This can be thought of as putting an object at the same initial position but with a different initial momentum. In this case, the system evolves along a separate energy contour. 

    \begin{figure}
    \centering
    \includegraphics[width=\linewidth]{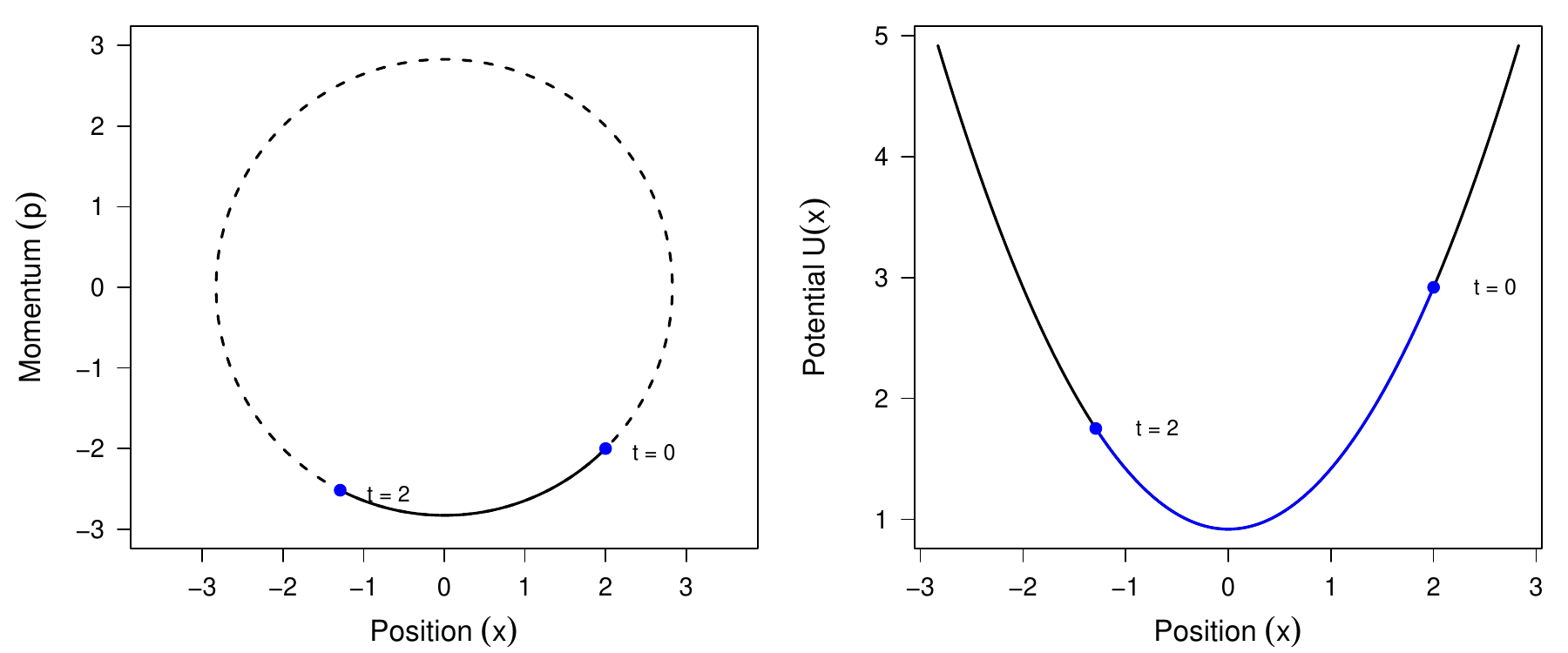}
    \caption{
    \text{Hamiltonian dynamics under a Gaussian potential.} The left panel depicts the trajectory of the particle in the $(x,p)$ phase space, starting from $(x_0, p_0) = (2, -2)$ and its state after $t = 2$ time units.
    The right panel shows the potential energy function $U(x) = -\log \pi(x)$, illustrating the quadratic bowl corresponding to $\pi \equiv N(0,1)$.}
    \label{fig:time_trajectory}
\end{figure}

\begin{figure}
    \centering
    \includegraphics[width=\linewidth]{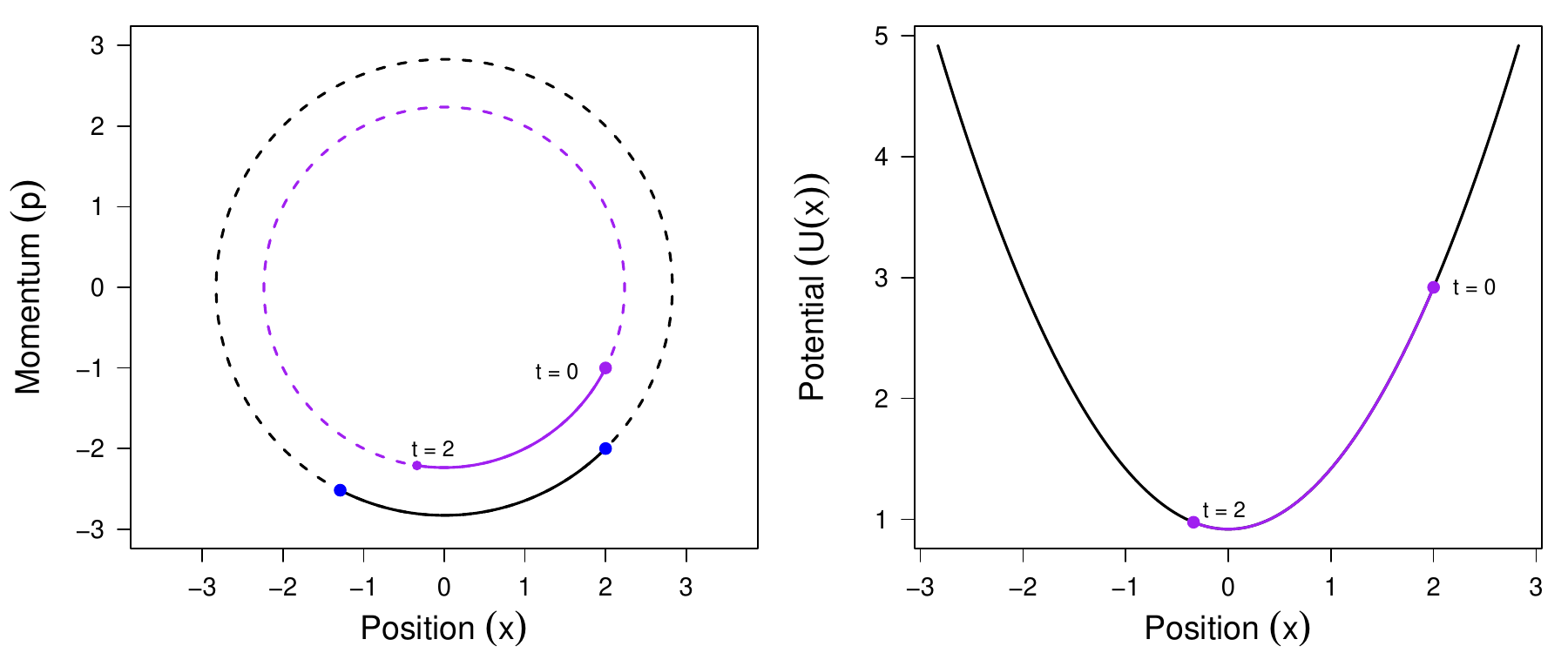}
    \caption{The left panel shows two distinct energy contours in the $(x,p)$ space corresponding to different initial momenta.  
    The right panel illustrates the particle’s position after two time units for the second trajectory. 
    }
    \label{fig:time_traj_diff_momentum}
\end{figure}
\end{runnexample} 

We now present the ideal HMC algorithm, building upon the theoretical foundation of Hamiltonian dynamics established so far. The ideal HMC is essentially an MHGJ algorithm with a specific choice of auxiliary variable and involution. Here, the joint target density  $\pi_{\text{Aug}}$ is defined as
\begin{equation}
    \pi_{\text{Aug}}(x,p) \propto e^{-H(x,p)} = e^{-U(x)} e^{-K(p)} \propto \pi(x) e^{-K(p)}\,.
\end{equation}
The Hamiltonian curves in the $(x,p)$ phase-space are the level curves of $H$, indicating points of equal value of the Hamiltonian; see the left plot of Figure~\ref{fig:time_traj_diff_momentum}. The ideal HMC sampler technically will sample from $\pi_{\text{Aug}}$, except the momentum samples are typically not saved. Notice here that by assumption, the marginal distribution of the momentum variable is $N(0, m)$, which is independent of the position. 

The next ingredient to build an MHGJ sampler is to identify the involution; this is where the Hamiltonian dynamics are employed. Let $T_s(x_t,p_t)$ denote the progression of the object after $s$ time units, starting from $(x_t, p_t)$. That is, let
\[
(x_{t+s}, p_{t+s}) = T_s(x_t,p_t)\,.
\]
The map $T_s$ has some nice properties. Namely,
\begin{enumerate}
    \item If $s = 0$, the particle stays where it is. That is, $T_0(x_t, p_t) = (x_t, p_t)$.
    \item The mapping is additive in time. That is, $T_{s_1 + s_2}(x_t, p_t) = T_{s_1} \left(T_{s_2} (x_t, p_t) \right) = T_{s_2} \left(T_{s_1} (x_t, p_t) \right)$.

    \item The inverse of $T_s$ is the mapping that goes back in time. That is, $T^{-1}_s(x_t, p_t) = T_{-s}(x_t, p_t)$.
\end{enumerate}
Despite these nice properties, the mapping $T_s$ itself is not an involution and must be used in conjunction with another map. Define the map that reverses the momentum of the object as
\[
i(x_t,p_t) := (x_t,-p_t)\,,
\]
called the momentum reversal map. Clearly, the mapping $i$ is an involution. When the kinetic energy function is chosen to be symmetric about zero, that is $K(p) = K(-p)$, then together, $T_s$ and $i$ satisfy time-reversal symmetry which states that flipping the momentum of a system, moving the object $s$ time units, and then flipping the momentum again is equivalent to going back in time $s$ units. That is,
\[
i \circ T_s \circ i (x_t,p_t) = T_{-s}(x_t, p_t)\,.
\]
Consider now the following composition map that pushes the object forward $s$ time units and then reverses the momentum:
\begin{equation}
    P_s(x_t,p_t) := i\circ T_s (x_t, p_t) = i\left(T_s(x_t,p_t) \right) = (x_{t+s}, -p_{t+s})\,.
\end{equation}
The mapping $P_s$ will be the final mapping that we will consider in the MHGJ algorithm. As it turns out, even though $T_s$ is not an involution, $P_s$ is. To see this, note that
\begin{align*}
    P_s\big(P_s(x_t,p_t)\big) & = i\big(T_s\big(i\big(T_s(x_t,p_t)\big)\big)\big) \\ 
& = T_{-s} \circ T_s(x_t,p_t) \\ 
& =  T_{s}^{-1} \circ T_s(x_t,p_t)\\ 
& = (x_t,p_t)\,.
\end{align*}
Thus, $P_s$ is an involution. Moreover, since $T_s$ is volume-preserving \citep[Liouville's theorem]{jordan2004steppingstones} and  the Jacobian of the momentum flip $i$ is a block diagonal matrix with determinant $\pm 1$,
therefore, 
\begin{equation}
\label{eq:determinant_Ps}
    \big|\det \nabla P_s(x,p)\big| = 1,
\end{equation}
showing that $P_s$ is volume-preserving as well. Intuitively, volume-preservation implies that the volume of a region in the $(x,p)$ space remains the same after $P_s$ acts on it. Finally, since $K(p)$ is an even function of $p$, we have
\[
H(x, -p) = U(x) + K(-p) = U(x) + K(p) = H(x,p)\,,
\]
ensuring energy invariance under momentum reversal. Using these ingredients, namely
\begin{itemize}
    \item $P_s$ is an involution,
    \item $P_s$ is volume-preserving, and
    \item $K(p)$ is symmetric around zero,
\end{itemize}
we can run an MHGJ algorithm, yielding an ideal HMC algorithm in Algorithm~\ref{alg:exact_hmc}. Specifically, Step~4 draws the auxiliary variable from its conditional distribution given the current iterate $x$; in this instance, the conditional distribution is independent of $x$. In Step~5, we apply the involution $P_s$ on $(x, p)$. Note here that this is akin to evolving the system using the Hamiltonian mapping $T_s$ and then reversing the momentum. Since $K(p) = K(-p)$, reversing the momentum does not change the outcome of the algorithm and thus is not necessary during implementation. Step~6 calculates the acceptance ratio, which is deterministically 1, since in Step~5 $(x^*,p^*)$ was obtained so as to conserve the Hamiltonian. Thus, the ideal HMC algorithm yields perfect acceptance and no rejections! Finally, since the momentum is resampled in Step~4, there is no need to save the momentum variables.

\begin{algorithm}
\caption{Ideal Hamiltonian Monte Carlo (HMC)}
\label{alg:exact_hmc}
\begin{algorithmic}[1]
\State \textbf{Initialize:} Set initial value $x^{(0)}$.
\For{$t = 0, 1, 2, \ldots, T-1$}
    \State Given current state $x^{(t)}$.
    \State Draw momentum $p \sim {\color{black}N}(0, m)$.
    \State Evolve the system for time $s$ to obtain
    $ (x^*, -p^*) = T_s(x^{(t)}, p).$
    \State Compute the Metropolis-Hastings ratio:
    \[
        r(x^{(t)}, x^*)
        =
        e^{-H(x^*, p^*) + H(x^{(t)}, p)}
        = 1.
    \]
    \State Set $x^{(t+1)} = x^*$.
\EndFor
\State \textbf{Return:} $\{x^{(t)}\}_{t=0}^{T-1}$.
\end{algorithmic}
\end{algorithm}

Perfect acceptance does not necessarily imply an excellent Markov chain. This is known for Gibbs samplers, which, although also having perfect acceptance, can exhibit slow mixing in various situations \citep{robe:sahu:1997}. The following running example will explain this point a bit more.

\begin{runnexample}
   We continue with the example of $\pi \equiv N(0,1)$ and $m = 1$ where recall
   \[
   H(x,p) = \dfrac{x^2}{2} + \dfrac{p^2}{2}\,.
   \]
   From \eqref{eq:gaussian_motion_eq}, we have
   \[
   T_s(x,p) = \begin{pmatrix}
       x \cos s + p \sin s \\ -x \sin s + p \cos s
   \end{pmatrix} \qquad \text{ and } \qquad P_s(x,p) = \begin{pmatrix}
       x \cos s + p \sin s \\ x \sin s - p \cos s
   \end{pmatrix}\,.
   \]
Using the above involution $P_s$, the ideal HMC in Algorithm~\ref{alg:exact_hmc} can be implemented. Below is an \texttt{R} function that implements the ideal HMC for the standard Gaussian target.
\begin{lstlisting}[language=R]
# ideal HMC function
normal_idealHMC <- function(s = 1, n = 1e4)
{
  x <- numeric(length = n)
  x[1] <- 0 # starting at zero position
  # don't need a starting value of p
  
  for(k in 2:n)
  {
    p <- rnorm(1)   # new momentum
    
    # simulating Hamiltonian dynamics forward s time units 
    # no practical need to flip momentum
    x[k] <- x[k-1]*cos(s) + p*sin(s)
    p <- -x[k-1]*sin(s) + p*cos(s)
  }
  return(x) # not returning momentum
}
\end{lstlisting}
 Figure \ref{fig:exact_normal} presents density estimates, trace plots, and autocorrelation function plots for three different choices of $s$. Noticeably, some choices of $s$ are better than others, and perfect acceptance does not guarantee excellent mixing of the chain. In this implementation, $s = 1,5$ yield well-behaved Markov chains, whereas $s = .1$ yields a significantly slower mixing Markov chain. This is because with $s = .1$, the object has not yet had the opportunity to travel far away from the current location, yielding smaller jump sizes, and thus higher autocorrelation. 
\begin{figure}
    \centering
    \includegraphics[width=1\linewidth]{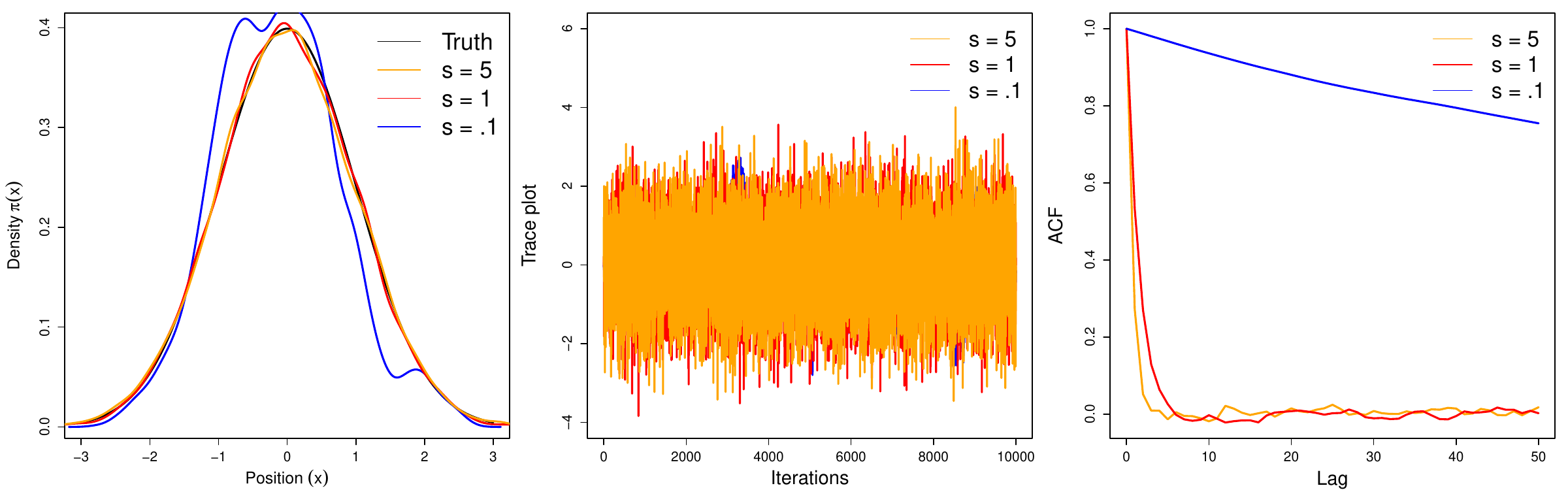}
    \caption{Density plots (left panel), trace plots (middle panel), and ACF plots (right panel) for ideal HMC on Gaussian target density with $s = 0.1, 1,5$.}
    \label{fig:exact_normal}
\end{figure}
\end{runnexample}

The ideal HMC algorithm is called as such since, for most realistic examples, an analytical expression for $T_s$ is not available. Thus, the involution cannot be applied directly. This roadblock leads to the widely accepted HMC algorithm, where the Hamiltonian dynamics are time-discretized to yield an approximation to $T_s$.

\subsection{Hamiltonian Monte Carlo algorithm}\label{sec:HMC_algo}

In practical implementations, the map $T_s$ is not available in closed-form, and Hamilton's equations must be approximated through numerical integration by discretizing time. This yields an approximation to $T_s$ that is used instead of $T_s$ in Algorithm~\ref{alg:exact_hmc} to obtain the proposal $x^*$. Recall Hamilton's equations
\begin{equation*}
   \dfrac{dp_t}{dt} = -\,\dfrac{\partial H(x_t,p_t)}{\partial x_t} = - \nabla U(x_t),
\qquad 
    \dfrac{dx_t}{dt} = \dfrac{\partial H(x_t,p_t)}{\partial p_t} = \nabla K(p_t) = \dfrac{p_t}{m}\,.
\end{equation*}
When a solution is analytically unavailable, a numerical integrator approximates $dt$ with an $\varepsilon > 0$. Such a discretization moves the object approximately according to Hamilton's equations for $s = {\color{black}\varepsilon}$ time units. As {\color{black} was} seen in Figure~\ref{fig:exact_normal}, too small an $s$ can yield a slow mixing Markov chain. Thus, such ${\color{black}\varepsilon}$-step discretizations are applied $L$ times so as to get $s = L\varepsilon $. However, any discretizer would only approximately be able to satisfy Hamilton's equations and thus would not be able to conserve the Hamiltonian exactly.  Consider a standard Euler discretization:
\begin{equation}\label{eq:euler_discretizer}
    \begin{aligned}
        p_{t + \varepsilon} &= p_{t} - \varepsilon\nabla U(x_t) \qquad \text{ and } \qquad 
        x_{t + \varepsilon} = x_{t} + \varepsilon\, \nabla K(p_{t + \varepsilon} ) = x_t + \varepsilon \color{black}{\dfrac{p_{t}}{m}}\,.
    \end{aligned}
\end{equation}
As ${\color{black}\varepsilon} \to 0$, consistent discretizers like the Euler discretization in \eqref{eq:euler_discretizer} will yield improved conservation of the Hamiltonian. However, this does not guarantee long-time stability under repeated composition. That is, repeated application of ${\color{black}\varepsilon}$-steps $L$ times worsens the quality of the approximation.
This has many consequences. First, any integrator will replace $T_s$ with an approximation, $T_{L, \varepsilon}$ in Algorithm~\ref{alg:exact_hmc}. In order for the resulting algorithm to still yield a $\pi$-invariant Markov chain, $T_{L, \varepsilon}$ and the resulting $P_{L, \varepsilon}$ must satisfy critical properties required by the MHGJ algorithm. That is, we require that
\begin{itemize}
    \item $P_{L, \varepsilon}$ is also an involution, and
    \item $P_{L, \varepsilon}$ is also volume preserving\,.
\end{itemize}
The standard Euler discretizer violates both these properties. However, a popular discretizer that guarantees both these properties is the leapfrog integrator.

\subsubsection{Leapfrog integrator}
\label{sec:leapfrog}

\begin{figure}
    \centering
    \includegraphics[width=\linewidth]{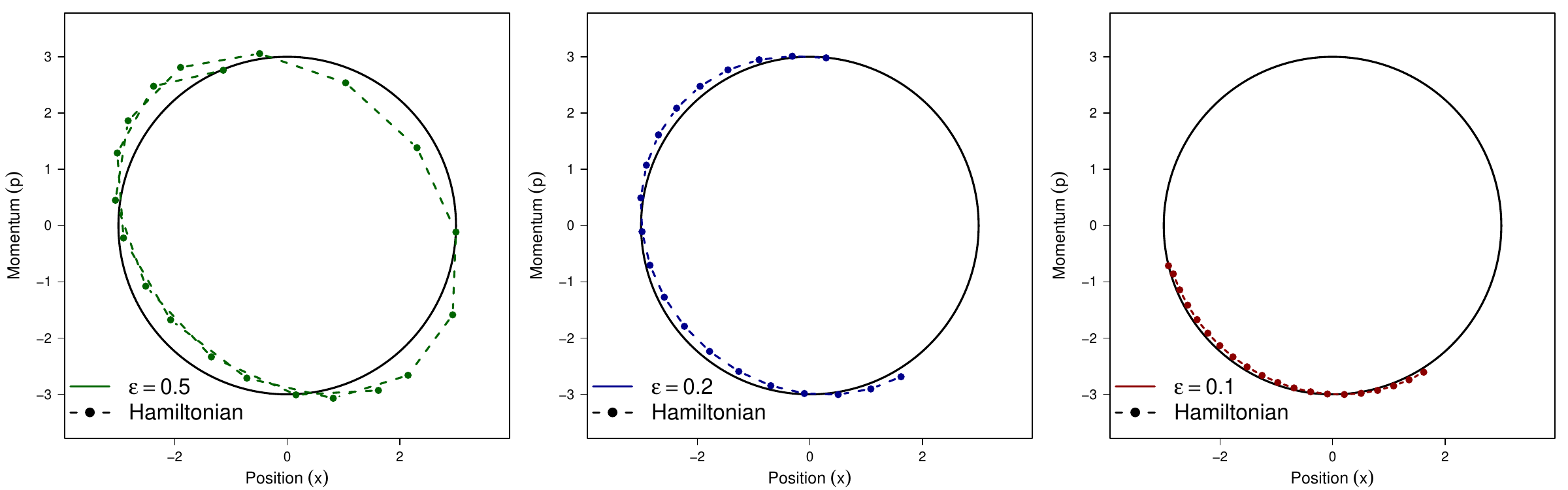}
    \caption{Approximated Hamiltonian paths with leapfrog discretization for different $\varepsilon = 0.5, 0.2, 0.1$ (from left to right) for standard Gaussian target distribution.}
    \label{fig:leapfrog_traj}
\end{figure}

The leapfrog integrator introduces a staggered update sequence in which the momentum $p$ is offset by $\varepsilon/2$ time units relative to the position $x$, yielding a scheme that forms the foundation of practical HMC implementations. Specifically, one jump of the leapfrog integrator first does an $\varepsilon/2$-Euler jump for $p$, followed by an $\varepsilon$-Euler jump of $x$ using the most updated state of $p$, with one last $\varepsilon/2$-Euler jump of $p$. We can write a single leapfrog update over one step of size $\varepsilon$ as:
\begin{enumerate}
    \item Half-step update for momentum:
    \begin{equation}
    p_{t + \frac{\varepsilon}{2}} = p_t - \dfrac{\varepsilon}{2} \, \nabla U(x_t)\,.
    \label{eq:first_step_momentum}
    \end{equation}
    \item Full-step update for position:
    \begin{equation}
    x_{t + \varepsilon} = x_{t} + \varepsilon \, \dfrac{p_{t + {\varepsilon/2}}}{m}\,.
   \label{eq:full_step_position}
    \end{equation}
    \item Final half-step update for momentum:
    \begin{equation}
    p_{t + \varepsilon} = p_{t + \frac{\varepsilon}{2}} - \dfrac{\varepsilon}{2} \, \nabla U(x_{t + \varepsilon})\,.
   \label{eq:final_step_momentum}
    \end{equation}
\end{enumerate}
This procedure approximately advances the system by $s = \varepsilon$ time units. To simulate a longer trajectory of duration $s = L \varepsilon$, one simply performs $L$ successive leapfrog steps, repeating \eqref{eq:first_step_momentum}–\eqref{eq:final_step_momentum}.  The final $(\varepsilon, L)$ leapfrog integrator steps are 
\begin{equation}\label{eq:L_leap-frog}
    \begin{aligned}
p_{t + \frac{\varepsilon}{2}} &= p_{t} - \frac{\varepsilon}{2} \nabla U(x_t), \\
x_{t + \varepsilon} &= x_{t} + \varepsilon \dfrac{p_{t + \varepsilon/2}}{m}, \\
p_{t + \frac{3\varepsilon}{2}} &= p_{t + \varepsilon/2} - \varepsilon \nabla U(x_{t + \varepsilon}), \\
 \vdots & \\
x_{t + L\varepsilon} &= x_{t + (L-1)\varepsilon} + \varepsilon\, \dfrac{p_{t + (2L-1)\varepsilon/2)}}{m}, \\
p_{t + L\varepsilon} &= p_{t + (2L-1)\varepsilon/2} - \frac{\varepsilon}{2} \nabla U(x_{t + L\varepsilon)})\,.
\end{aligned}
\end{equation}
The leapfrog integrator is a \textit{symplectic} integrator, implying both volume-preservation and the fact that repeated $\varepsilon$ steps do not yield poorer approximations. This is evidenced in Figure~\ref{fig:leapfrog_traj} where $L = 20$ repeated $\varepsilon$-steps do not yield a systematic propagation of error. {\color{black}Further}, it remains true that for $\varepsilon \approx 0$, the approximate path of the object stays closer to the true Hamiltonian than that for larger $\varepsilon$.

The leapfrog scheme in \eqref{eq:L_leap-frog} thereby defines a mapping $T_{L,\varepsilon}(x,p)$ that approximates the exact Hamiltonian flow $T_s(x,p)$, where $s = L\varepsilon$. As before, we define
\begin{equation}\label{eq:Ptilde}
P_{L,\varepsilon}(x,p) = i\big(T_{L,\varepsilon}(x,p)\big), \quad \text{where recall } i(x,p) = (x, -p).
\end{equation}
The function $P_{L,\varepsilon}$ in \eqref{eq:Ptilde} is both an involution and volume-preserving due to the nice properties of the leapfrog integrator. However, the resulting $P_{L, \varepsilon}(x,p)$ does not conserve energy exactly, and this implies that the acceptance ratio need not be 1. As before, flipping the sign is not important in practice. The resulting implementable version of HMC using the leapfrog integrator is provided in Algorithm~\ref{alg:lf_hmc}. This is the HMC algorithm, known and used widely in practice.
\begin{algorithm}
\caption{Hamiltonian Monte Carlo for $1$-dimensional targets}
\label{alg:lf_hmc}
\begin{algorithmic}[1]
\State \textbf{Initialize:} Set initial value $x^{(0)}$.
\For{$t = 0, 1, 2, \ldots, T-1$}
    \State Given current state $x^{(t)}$.
    \State Draw momentum $p \sim {\color{black}N}(0,m)$.
    \State Compute $(x^*, {\color{black}-}p^*) = {\color{black}T}_{L,\varepsilon}(x^{(t)}, p)$ using $L$ leapfrog steps of size $\varepsilon$.
    \State Compute the acceptance ratio:
    \[
        r(x^{(t)}, x^*)
        =
        \exp\{-H(x^*, p^*) + H(x^{(t)}, p)\}.
    \]
    \State Draw $W \sim \text{Uniform}(0,1)$, independently.
    \If{$W \leq \min\{1, r(x^{(t)}, x^*)\}$}
        \State Set $x^{(t+1)} = x^*$.
    \Else
        \State Set $x^{(t+1)} = x^{(t)}$.
    \EndIf
\EndFor
\State \textbf{Return:} $\{x^{(t)}\}_{t=0}^{T-1}$.
\end{algorithmic}
\end{algorithm}

\begin{runnexample}
    For $\pi \equiv N(0,1)$ and $m = 1$,
    \[
    \nabla K(p) = p \text{ and } \nabla U(x) = x\,.
    \]
We implement Algorithm~\ref{alg:lf_hmc} in the \texttt{R} code below.
\begin{lstlisting}[language=R]
normal_HMC <- function(L = 10, eps = .1, n = 1e4)
{
  x <- numeric(length = n)
  x[1] <- 0 # starting from x = 0
  # vectorizing this outside to save time in R
  momentums <- rnorm(n) 
  
  for(k in 2:n)
  {
    p <- momentums[k] # new momentum
    x_prop <- x[k-1]
    
    # one half-Euler step
    p_prop <- p - eps/2 * x_prop  
    # let the frog leap!
    for(l in 1:L)
    {
      x_prop <- x_prop + eps * p_prop
      if(l != L) p_prop <- p_prop - eps*x_prop 
    }
    # one last half-Euler step
    p_prop <- p_prop - eps/2 * x_prop
    
    # Accept-reject on log scale
    log.ratio <-  (-x_prop^2 - p_prop^2 + x[k-1]^2 + p^2)/2
    if(log(runif(1)) < log.ratio)
    {
      x[k] <- x_prop
    } else{
      x[k] <- x[k-1]
    }
  }
  return(x)
}    
\end{lstlisting}
We call the function with $s = 1$ for various choices of $\varepsilon$ and present the results in Figure~\ref{fig:lfd_with_s1}. The quality of the samples is fairly close to that of the ones obtained from the ideal HMC.
\begin{figure}
    \centering
    \includegraphics[width=1\linewidth]{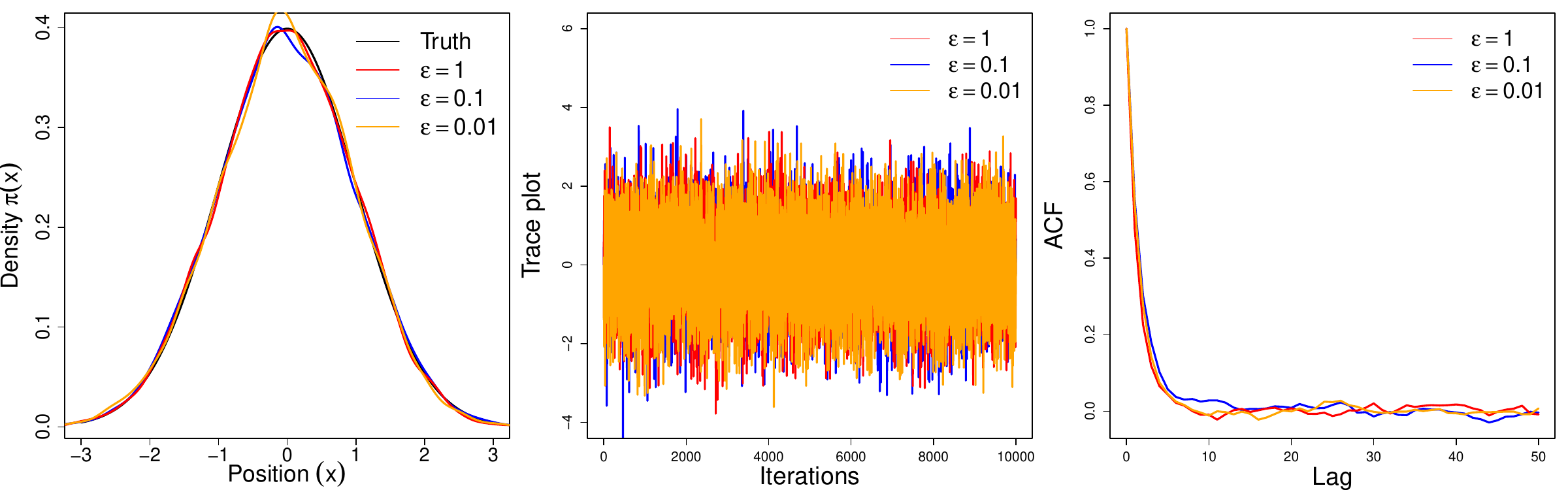}
    \caption{Density plots (left panel), trace plots (middle panel), and ACF plots (right panel) {\color{black}of} HMC on Gaussian target density with different $\varepsilon = 0.01, 0.1, 1$ with $s = L\varepsilon = 1.$}
    \label{fig:lfd_with_s1}
\end{figure}
\end{runnexample}

\subsubsection{HMC in higher dimensions}
\label{sub:hmc_higher_dim}

So far, we have inherently assumed that $\pi$ is defined on a 1-dimensional space. Naturally, this is rarely the case for the problems in which MCMC is employed. The extension to a general $d$-dimensional space is fairly straightforward. The potential $U(x)$ remains $U(x) = -\log \pi(x)$. However, the auxiliary variable (momentum) is also $d$-dimensional and for an appropriately chosen $p \times p$ mass matrix, $M$, the kinetic energy considered is
\[
K(p) = \frac{1}{2} p^\top M^{-1} p\,.
\]
Thus, the Hamiltonian on $\mbb{R}^{2d}$ takes the form 
\[
H(x,p) = U(x) + K(p)
       = -\log \pi(x) + \frac{1}{2} p^\top M^{-1} p\, .
\]
Consequently, $\nabla K(p) = M^{-1} p$. Given a step size $\varepsilon > 0$ and a number of steps $L$, the leapfrog in general dimensions can be written down as an extension of \eqref{eq:first_step_momentum}--\eqref{eq:final_step_momentum} as follows:
\begin{equation}\label{eq:L_leap-frog-d}
    \begin{aligned}
p_{t + \frac{\varepsilon}{2}} &= p_{t} - \frac{\varepsilon}{2} \nabla U(x_t), \\
x_{t + \varepsilon} &= x_{t} + \varepsilon M^{-1} p_{t + \varepsilon/2}, \\
p_{t + \frac{3\varepsilon}{2}} &= p_{t + \varepsilon/2} - \varepsilon \nabla U(x_{t + \varepsilon}), \\
 \vdots & \\
x_{t + L\varepsilon} &= x_{t + (L-1)\varepsilon} + \varepsilon\, M^{-1} p_{t + (2L-1)\varepsilon/2)}, \\
p_{t + L\varepsilon} &= p_{t + (2L-1)\varepsilon/2} - \frac{\varepsilon}{2} \nabla U(x_{t + L\varepsilon)})\,.
\end{aligned}
\end{equation}
The final HMC algorithm in general dimensions is provided in Algorithm~\ref{alg:lf_hmc_d_dim}.

\begin{algorithm}
\caption{Hamiltonian Monte Carlo for $d$-dimensional targets}
\label{alg:lf_hmc_d_dim}
\begin{algorithmic}[1]
\State \textbf{Initialize:} Set initial value $x^{(0)}$.
\For{$t = 0, 1, 2, \ldots, T-1$}
    \State Given current state $x^{(t)}$.
    \State Draw momentum $p \sim {\color{black}N}_d(0,M)$.
    \State Compute $(x^*, {\color{black}-}p^*) = {\color{black}T}_{L,\varepsilon}(x^{(t)}, p)$ using $L$ leapfrog steps of size $\varepsilon$ using \eqref{eq:L_leap-frog-d}.
    \State Compute the acceptance ratio:
    \[
        r(x^{(t)}, x^*)
        =
        \exp\{-H(x^*, p^*) + H(x^{(t)}, p)\}.
    \]
    \State Draw $W \sim \text{Uniform}(0,1)$, independently.
    \If{$W \leq \min\{1, r(x^{(t)}, x^*)\}$}
        \State Set $x^{(t+1)} = x^*$.
    \Else
        \State Set $x^{(t+1)} = x^{(t)}$.
    \EndIf
\EndFor
\State \textbf{Return:} $\{x^{(t)}\}_{t=0}^{T-1}$.
\end{algorithmic}
\end{algorithm}

\section{Choosing tuning parameters}\label{sec:optimal_tuning}

A successful implementation of an HMC algorithm requires careful tuning of $s, \varepsilon, L,$ and $M$. In this section, we provide recommendations for these parameters based on both heuristics and theory. Typically, a good strategy is to first choose $M$ according to Section~\ref{sec:mass}. Next, fix a small $L$ and choose  $\varepsilon$ according to the guidelines in Section~\ref{sec:epsilon}. Finally, using $M$ and $\varepsilon$ thus chosen, choose $L$ and consequently $s = L \varepsilon$ based on the guidelines in Section~\ref{sec:number_lf}.

\subsection{Mass matrix $M$}
\label{sec:mass}

A first practical and natural choice for $M$ is setting $M = \mathbb{I}_d$. Usually, a practitioner first tries this choice, and if the sampler is ineffective, a more informed choice of $M$ is made. A practical workflow is demonstrated in Section~\ref{sec:data_analysis}. When a mass matrix different from $\mathbb{I}_d$ is employed, this is akin to linear preconditioning.

The choice of the mass matrix controls how the algorithm moves through the augmented parameter space. Intuitively, the mass matrix determines the relative speed at which the $d$ different directions are explored during the simulated Hamiltonian trajectories. If these speeds are poorly matched to the scale of the target distribution, the resulting trajectories can become inefficient, requiring very small step sizes and leading to poor mixing. From a geometric perspective, a linear transformation of the position variables is equivalent to the inverse linear transformation of the momentum variables, and the mass matrix provides a convenient way to implement this re-scaling without explicitly reparametrizing the model. Considering this, an ideal choice of $M$ is such that 
\[
M^{-1}  =  \text{Cov}_{\pi}(X) =: \Sigma\,.
\]
However, $\Sigma$ will typically be unavailable in practice, in which case, a warm-up chain is run from which $\Sigma$ is estimated using a sample covariance matrix employing the sample mean, $\bar{x} = T^{-1} \sum_{t=0}^{T-1}x^{(t)}$:
\begin{equation}\label{eq:inv_cov_mat}
    \widehat{\Sigma} = \dfrac{1}{T - 1}\sum_{t=0}^{T-1} (x^{(t)} - \bar{x})(x^{(t)} - \bar{x})^{\top}\,.
\end{equation}
Consequently, $M = \widehat{\Sigma}^{-1}$ is chosen. There are two disadvantages of this method, (i) the estimator in \eqref{eq:inv_cov_mat} may be poor due to poor warm-up samples, and (ii) the sampler can be expensive to implement due to the need of employing a matrix-vector multiplication for every leapfrog step. 

A reasonable alternative is to employ a diagonal $M$ with elements $1/\textrm{diag}\left(\widehat{\Sigma} \right)$. That is, the diagonals of $M$ are set to be the marginal inverse sample variances for all components. This is a computationally cheaper alternative that can often work well, which we also employ in Section~\ref{sec:data_analysis}. However, recently, \cite{hird:2025} show that it is in fact possible for this type of preconditioning to yield worse results than no preconditioning at all. They also propose a computationally viable alternate preconditioning method that proves to be more effective than diagonal preconditioning.

While an appropriate mass matrix can dramatically improve performance, it is important to recognize the limitations of linear preconditioning. Mass matrix adaptation is most effective when the target is well approximated by an elliptically contoured distribution, such as a multivariate Gaussian. For more complex targets exhibiting strong nonlinearity, varying curvature, or pronounced non-elliptical structure, no fixed linear transformation can simultaneously regularize the geometry everywhere. In such cases, \cite{girolami:2011} propose Riemannian manifold HMC methods that incorporate position-dependent geometry where the mass matrix $M$ is allowed to also depend on $x$. We expand on this a bit more in Section~\ref{sec:Riemannian_HMC}.

\subsection{Discretization size $\varepsilon$}
\label{sec:epsilon}

Having found an appropriate $M$, we move on to choosing $\varepsilon$. This parameter is more easily tunable due to the measurable effects and consequences of choosing a particular $\varepsilon$.

Recall that the acceptance probability for the ideal HMC in Algorithm~\ref{alg:exact_hmc} is exactly 1 due to the exact conservation of the Hamiltonian. In most situations, the solution to Hamilton's equations is unavailable, and the leapfrog integrator in Section~\ref{sec:leapfrog} is used to simulate the Hamiltonian dynamics. This approximation leads to a change in the Hamiltonian, no longer ensuring an acceptance probability of exactly 1. Recall, however, that the Hamiltonian is increasingly better conserved as $\varepsilon \to 0$. Thus, the choice of the discretization step, $\varepsilon$, dictates the acceptance probability of the algorithm.  A smaller value of $\varepsilon$ will yield higher acceptance, and a larger value of $\varepsilon$ will yield lower acceptance. Further, due to the characteristics of the leapfrog integrator, for a fixed $\varepsilon$, the acceptance probability is minimally affected by the choice of $L$.

\begin{runnexample}
For our running example, we implement the HMC algorithm for $s = L \varepsilon = 1$ and $s = L \varepsilon = 10$ in Table~\ref{tab:hmc_two_columns}. We make two observations: first, the same value of $\varepsilon$ for different values of $L$ yields similar acceptances. This indicates that indeed the acceptance probability is primarily affected by $\varepsilon$.  Second, acceptances decrease as $\varepsilon$ increases. 
    \begin{table}
\centering
\caption{Acceptance rates for Leapfrog-based HMC algorithm with varying $(L, \varepsilon)$ settings. Total trajectory length is $s = L \varepsilon$.}
\label{tab:hmc_two_columns}
\renewcommand{\arraystretch}{1.2}
\begin{tabular}{c c}
\begin{tabular}{c c c c}
\hline
\multicolumn{3}{c}{$s = L \varepsilon = 1$} \\
\hline
Chain & $\varepsilon$ & $L$ & Acceptance \\
\hline
1 & 0.01 & 100 &  1.0000 \\
2 & 0.1  & 10 & 0.9996 \\
3 &  1   & 1 & 0.9181 \\
\hline
\end{tabular}
&
\begin{tabular}{c c c c }
\hline
\multicolumn{3}{c}{$s = L \varepsilon = 10$} \\
\hline
Chain & $\varepsilon$ & $L$ &Acceptance \\
\hline
1 & 0.1  & 100 & 0.9995 \\
2 &  1   & 10 & 0.9155 \\
3 & 10   & 1 & 0.0037 \\
\hline
\end{tabular}
\end{tabular}
\end{table}
As always, large acceptance is not necessarily an indicator of a good Markov chain. In Figure~\ref{fig:lfd_with_s10}, we present visual summaries of the HMC algorithm with $s = L \varepsilon = 10$. This figure is in stark contrast to Figure~\ref{fig:lfd_with_s1}, where all choices of $\varepsilon$ yielded well-behaved Markov chains. Here, when $\varepsilon$ is too large, the chain gets stuck at the current position for a while since the Hamiltonian of the proposal deviates significantly from the Hamiltonian of the current value. 
\begin{figure}
    \centering
    \includegraphics[width=1\linewidth]{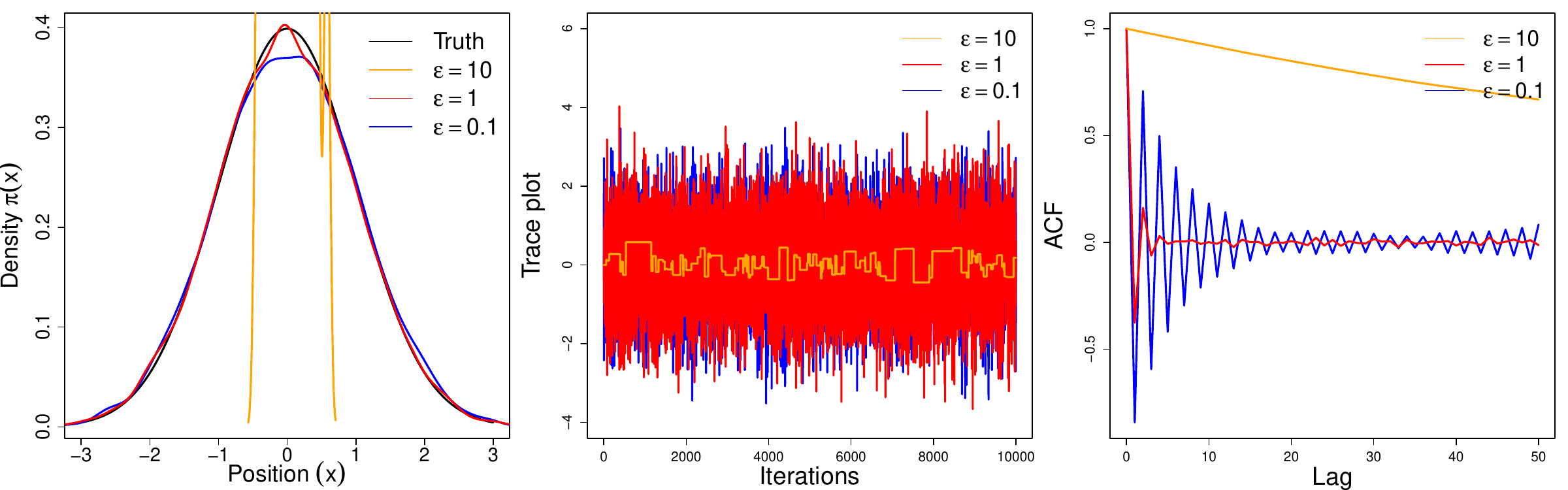}
    \caption{Density plots (left panel) and trace plots (right panel) for exact HMC on Gaussian target density with different $\varepsilon = 0.1, 1, 10$ with $s = L\varepsilon = 10$.}
    \label{fig:lfd_with_s10}
\end{figure}
\end{runnexample}

As discussed previously, a larger acceptance probability need not imply an excellent Markov chain. Using techniques from the optimal scaling literature \citep{roberts2001optimal}, \cite{beskos2013optimal} showed that as dimension $d \to \infty$, a user should tune $\varepsilon$ to yield an acceptance rate of $0.651$. However, for large but finite $d$, their simulations showed that higher acceptance rates up to $0.85$ can yield better results. In general, users often choose $\varepsilon$ aiming for $65\%$ acceptance for complicated and high-dimensional target distributions and tune for acceptances closer to $85\%$ for simpler targets. The tuning of $\varepsilon$ can then be done using shorter length Markov chains than the final run, employing a starting value that is generally in an area of high probability.

\subsection{Number of leapfrog steps and $s$}
\label{sec:number_lf}

Having chosen $M$ and $\varepsilon$, we move on to choosing the number of leapfrog steps, $L$. The performance of HMC depends critically on the number of leapfrog steps. Too few steps and the chain will explore the space like a slow random walk; too many, and we waste computation. A reasonable way then to choose $L$ is to assess the computational budget. Each step of the leapfrog requires the calculation of a gradient vector, which is often the leading cost in the algorithm. Thus, $L$ can be chosen based on the computation available. Note that for a given $\varepsilon$, a large $L$ is unlikely to affect the acceptance probability of the HMC algorithm. Thus, if we have the computational bandwidth to increase $L$, it is often preferable.

There is, however, a limit to how large we might want $L$ to be. It is possible for $L$ to be large enough that a push of $s = L \varepsilon$ time units brings the object back close to the original position. This depends on the period of the Hamiltonian. Ideally, we would like to choose $L$ so that $s = L \varepsilon$ is close to the half period of the underlying Hamiltonian. Then, the proposal $x^*$ is in general far from the current state of the object. This is challenging to do in practice since different parts of the state-space can have different periodicity. We discuss a popular adaptive procedure that attempts to solve this problem in Section~\ref{sec:NUTS}.

\begin{runnexample}
In Figure~\ref{fig:traj_period}, we demonstrate the quarter, half, and full period of the Gaussian Hamiltonian. For the same choice of $\varepsilon$ (and thus same acceptance), each choice of $L$ will yield different quality Markov chains. Increasing $L$ beyond a value of $31$ will be both wasteful and statistically inefficient.
\begin{figure}
    \centering
    \includegraphics[width=\linewidth]{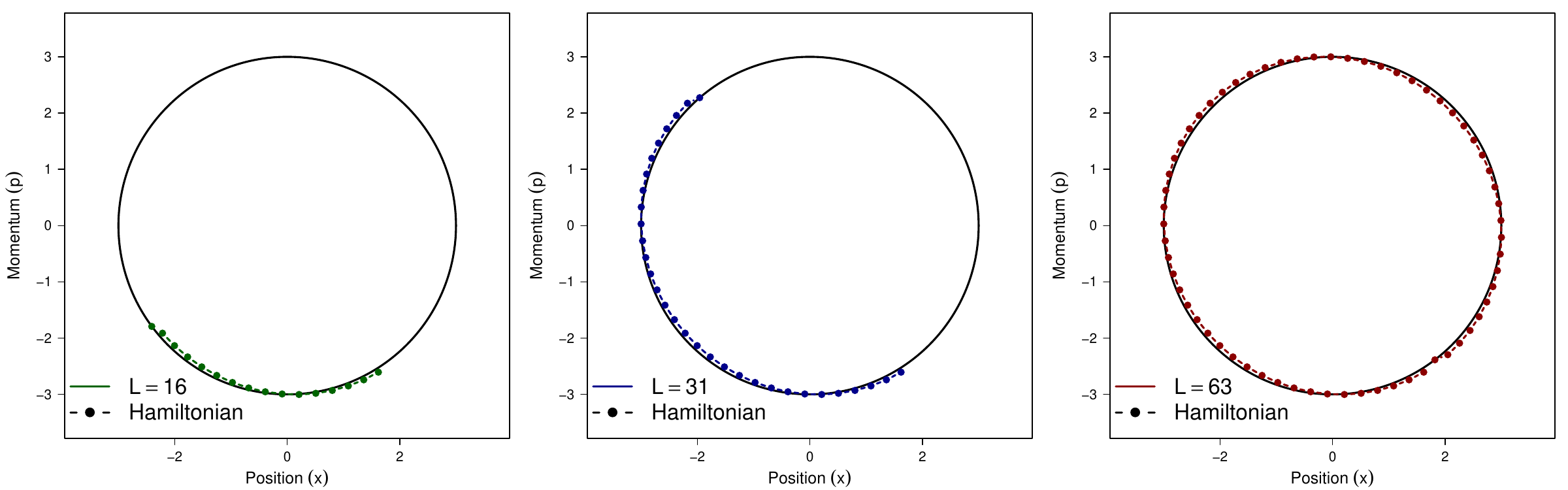}
    \caption{Approximated Hamiltonian paths with leapfrog discretization for fixed $\varepsilon = 0.1$ and with different $L$ with time-trajectories $s= L\varepsilon$ at quarter period (left), half-period (middle), full-period (right) for standard Gaussian target distribution.}
    \label{fig:traj_period}
\end{figure}

\end{runnexample}

In the next section, we discuss the implementation of HMC for a Bayesian logistic regression model on a real and popular dataset, demonstrating some key learnings from this section.

\section{Example implementation}\label{sec:data_analysis}

We present an HMC implementation for Bayesian logistic regression on the popular Pima Indian diabetes dataset \citep{chopin2017leave}. This dataset was originally collected by the National Institute of Diabetes and Digestive and Kidney Diseases and contains information on adult female individuals of Pima Indian heritage, all aged at least $21$ years and living near Phoenix, USA. Each response is a binary variable indicating the presence or absence of diabetes, together with covariates derived from medical examinations.
We analyze a subset of the data available in the \texttt{MASS} package in \texttt{R}, consisting of $n = 200$ observations with no missing values. \textcolor{black}{There are $d = 8$ covariates, including the intercept, number of pregnancies} (\texttt{pregnant}), plasma glucose concentration measured two hours after an oral glucose tolerance test (\texttt{glucose}), diastolic blood pressure in mmHg (\texttt{pressure}), triceps skin-fold thickness in mm (\texttt{triceps}), body mass index (\texttt{mass}), defined as weight in kilograms divided by the square of height in meters, and the diabetes pedigree function (\texttt{pedigree}), which summarizes the genetic predisposition based on family history. 

To study the relationship between these covariates and diabetes status, we employ a Bayesian logistic regression model of the form
\[
Y_i \overset{\text{ind}}{\sim} \mathrm{Bernoulli}(p_i) \;\text{for all}\; i = 1,2, \ldots, n.
\]
Denoting $z_{i}$ as the covariate for the $i$th observation and the corresponding regression coefficient vector $\beta$, we model each $p_i$ with the logit linked linear covariates as 
\[
\log{\left(\dfrac{p_i}{1-p_i}\right)} = z_{i}^{\top}\beta\,,
\]
with independent priors on the regression coefficient $\beta \sim {\color{black}N}_d(0,\sigma^2 \mathbb{I}_d)$. We choose the hyperparameter $\sigma^2 = 100$ to impose weak prior information. The resulting posterior is proper but intractable, and thus MCMC is used to obtain samples for posterior summaries.

\begin{figure}
    \centering
    \includegraphics[width=0.8\linewidth]{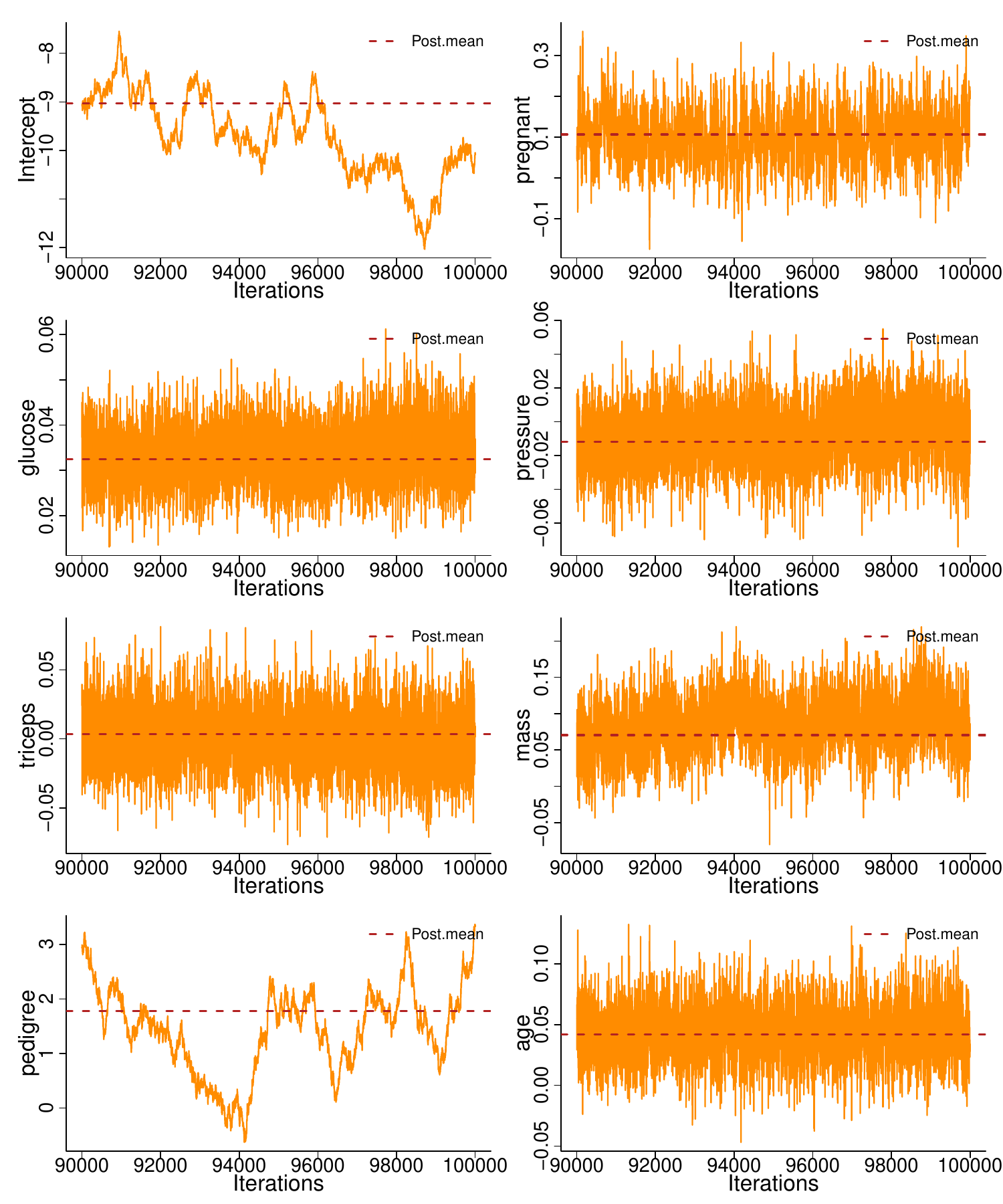}
    \caption{Traceplots of components of $\beta$ with posterior mean (in dotted line) of the resulting Markov chain from the naive HMC sampler. The figure indicates that, except for \texttt{Intercept} and \texttt{pedigree}, all components are fast mixing.}
    \label{fig:traceplots_wo_pre}
\end{figure}

\begin{figure}
    \centering
    \includegraphics[width=0.8\linewidth]{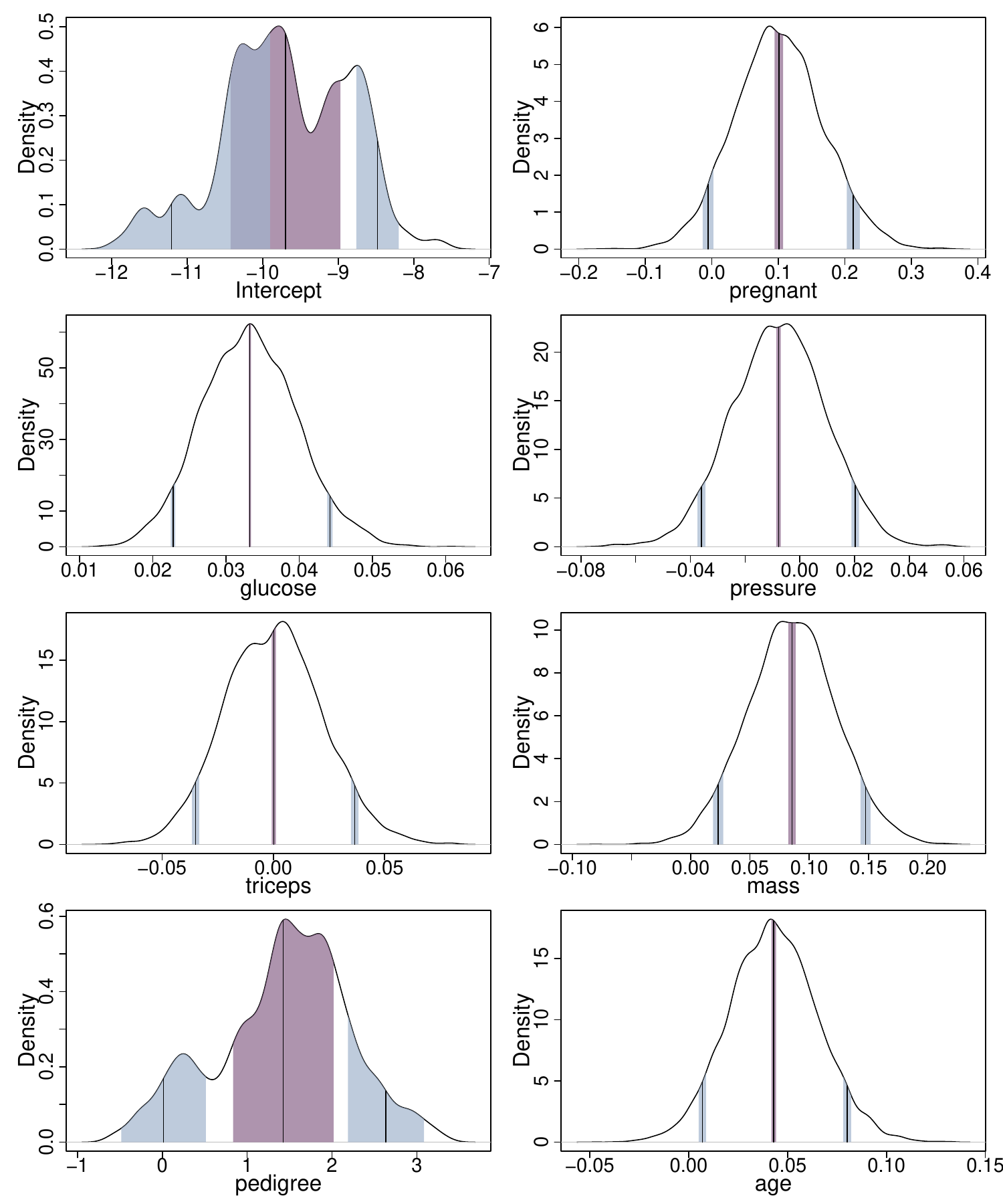}
    \caption{Density estimates for the components of $\beta$, with posterior means and the $5\%$ and $95\%$ estimated posterior quantiles from the naive HMC sampler. Bands around vertical lines show Monte Carlo uncertainty. Plot is made using \texttt{R} package \texttt{SimTools}.}
    \label{fig:densityplots_wo_pre}
\end{figure}

First, we illustrate the importance of employing a suitable mass matrix as described in Section~\ref{sec:mass}. We begin by running a baseline HMC algorithm using the identity mass matrix $M = \mathbb{I}_d$, following Algorithm~\ref{alg:lf_hmc_d_dim}. The sampler is run for $10^5$ iterations with tuning parameters $L = 20$ and $\varepsilon = 2.1 \times 10^{-5}$, chosen in accordance with the guidelines in Section~\ref{sec:optimal_tuning}.  The resulting trace plots shown in Figure~\ref{fig:traceplots_wo_pre} reveal that while most components of $\beta$ exhibit satisfactory mixing, two components \texttt{Intercept} and \texttt{pedigree} mix slowly due to the variability of different scales present in the posterior. This lack of mixing leads to unreliable posterior density estimates, as displayed in Figure~\ref{fig:densityplots_wo_pre}.


\begin{figure}
    \centering
    \includegraphics[width=0.8\linewidth]{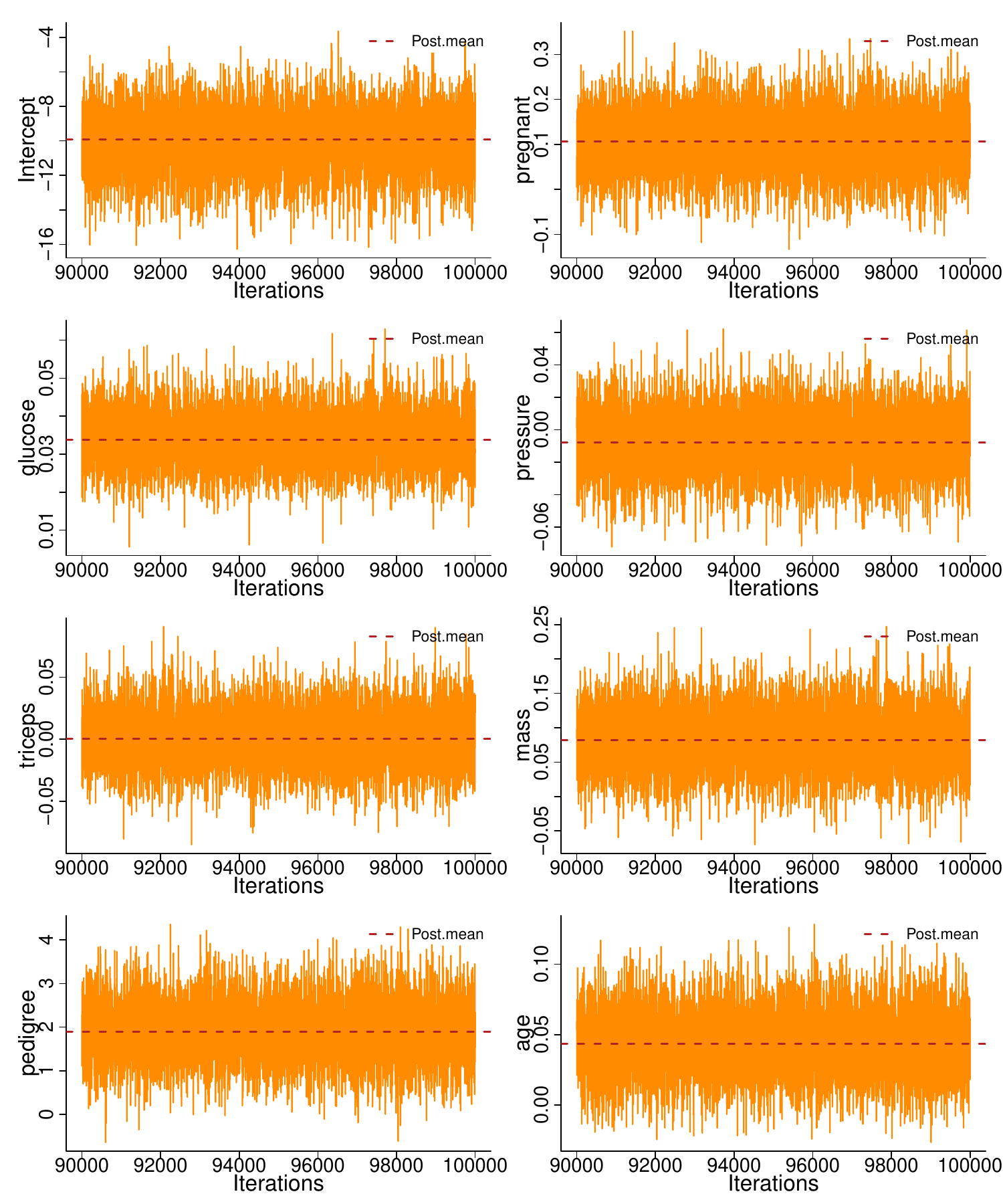}
    \caption{Trace plots of components of $\beta$ with posterior mean (in dotted line).}
    \label{fig:traceplots}
\end{figure}

\begin{figure}
    \centering
    \includegraphics[width=0.8\linewidth]{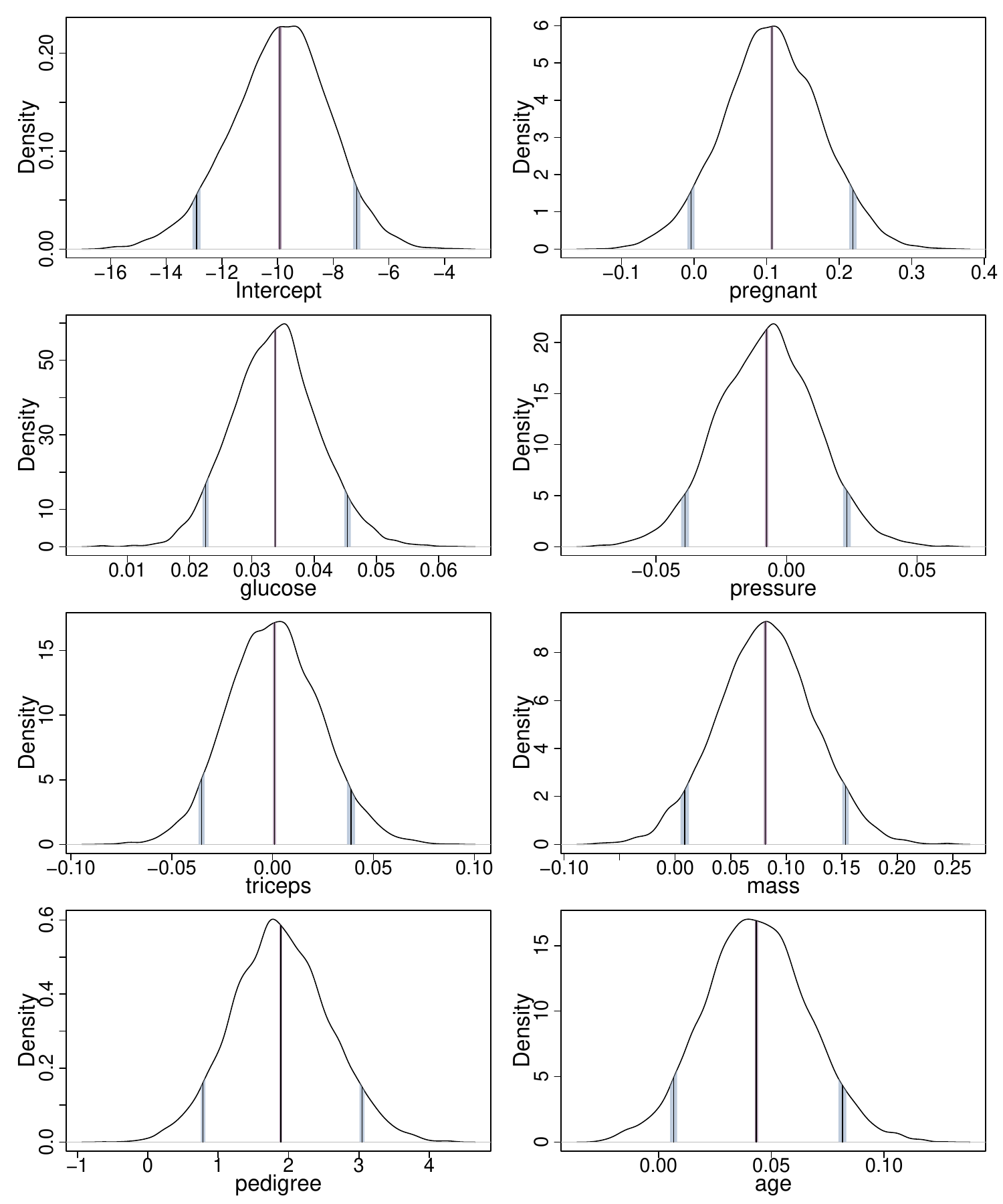}
   \caption{Density estimates for the components of $\beta$, with posterior means and the $5\%$ and $95\%$ estimated posterior quantiles for pre-conditioned sampler. The figure indicates that all components are well estimated.}
    \label{fig:densityplots}
\end{figure}

To mitigate the scale imbalance, we employ the diagonal preconditioning strategy described in Section~\ref{sec:mass}. Specifically, we use a computationally cheaper diagonal mass matrix $M$, comprised of the inverse marginal sample variances estimated from warm-up iterations of a naive HMC run. We then rerun the HMC sampler for $10^5$ iterations using the same leapfrog length $L = 20$, but with a substantially larger step size $\varepsilon = 0.11$, which maintains an acceptance rate consistent with the tuning considerations of Section~\ref{sec:optimal_tuning}. Figure~\ref{fig:traceplots} displays trace plots for the final $10^4$ iterations of this preconditioned sampler, and the corresponding posterior density estimates are shown in Figure~\ref{fig:densityplots}. Together, these results demonstrate a significant improvement in mixing across all components of $\beta$, as well as substantially more reliable posterior density estimation. The \texttt{R} codes to reproduce the simulation studies and real data analysis are available on GitHub\footnote{Available at the repository \url{https://github.com/ArghyaStat/HMC_for_dummies}.}.

\section{HMC variants}\label{sec:HMC_variants}

The potential power of HMC algorithms, as well as the need for its careful tuning, has led to several extensions and variations. We provide a short review of popular variants of the HMC algorithm proposed in the literature.

\subsection{NUTS}\label{sec:NUTS}

Perhaps the most famous and powerful extension of the HMC algorithm is the No-U-Turn Sampler (NUTS). As seen in our previous sections, the efficiency of HMC critically depends on the tuning of the step size $\varepsilon$ and the number of leapfrog steps $L$. If $\varepsilon$ is too large relative to the local curvature of $\pi$, the numerical trajectory can diverge, while an excessively small $\varepsilon$ leads to insufficient exploration. The NUTS  \citep{hoffman:2014} eliminates the need to pre-specify the integration time $s = L \varepsilon$ by adaptively determining how long to simulate Hamiltonian dynamics. NUTS recursively constructs a trajectory by simulating Hamiltonian dynamics in both forward and backward time directions uniformly at random. The orbit is thus extended in that direction by doubling its length. The specifics of this ensure that the trajectory does not double back on itself. The resulting set of candidate states forms an orbit that is both symmetric and volume-preserving, thereby maintaining the reversibility of the Markov transition. From the constructed orbit, a single proposal $(x^*, p^*)$ is drawn using a sampling scheme that preserves invariance. This adaptive construction enables NUTS to automatically adjust the integration time according to the local geometric features of the target distribution, substantially improving sampling efficiency without requiring manual tuning of $L$. Stan \citep{stan:dev:team:2017} provides a widely used implementation of NUTS.

While NUTS adaptively addresses the integration length, it still employs a single global step size $\varepsilon$ across the entire sampling process. {\color{black} In Stan, this choice of $\varepsilon$ is determined in the warm-up phase using dual averaging \citep{nesterov2009primal} targeting  desired acceptance rate.} In multiscale problems, regions of high curvature in $U(x)$ may demand smaller step sizes for stability, whereas flatter regions permit larger ones. A fixed $\varepsilon$, therefore, presents a trade-off between numerical stability and efficiency. \cite{bou:nuts:gibbs:2025} recently introduce orbit-level adaptive step-size variants of NUTS, which select $\varepsilon$ separately for each orbit while maintaining detailed balance. Building further on this idea, \cite{bou:walnuts:2025} propose the WALNUTS (Within-orbit Adaptive Leapfrog No-U-Turn Sampler), where the step size $\varepsilon_t$ is allowed to vary within each orbit as a function of the local curvature, thereby ensuring stability in regions of high curvature while preserving efficiency in flatter regions. This localized adaptation improves robustness and energy control without violating reversibility or detailed balance. 

\subsection{Stochastic gradient HMC}\label{sec:Stoc_grad_hmc}
One of the crucial bottlenecks of HMC is the need to compute the gradient of the log target repeatedly within a single iteration. Modern models frequently deal with 
datasets having a massive number of observations. In these ever-more-common scenarios of massive data, full gradient computation is infeasible. For example, suppose we want to sample from the posterior distribution of $\theta$, the model parameter given a set of independent observations $\mc{D} = \{y_1, \ldots, y_n\}$, denoted by
\begin{equation}
    \pi(\theta | \mc{D}) = \left\{\D\prod_{j=1}^{n} \log{p(y_j | \theta)} \right\}p(\theta) \propto \exp{\left(-U(\theta) \right)}\,,
\end{equation}
where $\log{p(y_j | \theta)}$ is the likelihood of $\theta$ based on $y_j$ and $p(\theta)$ is the prior density of $\theta$. The gradient of the potential energy function is thus given by
\begin{equation}
    \nabla U(\theta) = - \D\sum_{j=1}^{n} \nabla \log{p(y_j | \theta)} - \nabla p(\theta)\,.
\end{equation}
Traditional stochastic methods \citep{well:teh:2011} enjoy success by approximating $\nabla U(\theta)$ by $\nabla \widetilde{U}(\theta)$  using minibatches of data, say $\widetilde{\mc{D}}$, of size $k \ll n$ to scale algorithms for large $n$ as
\begin{equation}
    \nabla \widetilde{U}(\theta) = - \D\dfrac{k}{n}\sum_{y_j : y_j \in \widetilde{\mc{D}}} \nabla \log{p(y_j | \theta)} - \nabla p(\theta)\,.
\end{equation}
\cite{chen:2014} propose variants of stochastic gradient HMC with replacing $\nabla U$ with $\nabla \tilde{U}$. But HMC with stochastic gradient $\nabla \tilde{U}$ can remain costly due to the accept-reject step requiring the need the evaluate the full Hamiltonian. \cite{chen:2014} introduces a friction term in the Hamiltonian dynamics which allows the proposed sample path to be used without an accept-reject step. Note that the use of stochastic gradients comes at the cost of exactness of the algorithm.

 \cite{ma:2017} provide a complete framework for stochastic gradient MCMC methods. They discussed a unified underlying stochastic dynamical system where HMC and SGHMC are special cases. \cite{zou:2021} propose a stochastic variance-reduced HMC method for sampling from a smooth and strongly log-concave distribution. Their algorithm exhibits a faster rate of convergence and better gradient complexity than vanilla HMC and stochastic gradient HMC methods across a wide range of regimes. For a complete review of stochastic gradient MCMC methods, see \cite{nemeth2021stochastic}.

\subsection{Riemannian manifold HMC}\label{sec:Riemannian_HMC}

Sampling efficiency of HMC highly depends on the local geometric structure of the target distribution. Several extensions of HMC have sought to exploit this structure by modifying the kinetic energy function $K(p)$ through the introduction of global preconditioning matrices. A major advancement in this direction was proposed by \citet{girolami:2011}, who extended HMC from Euclidean manifolds to Riemannian manifolds by defining a position-dependent Gaussian momentum distribution. This is referred to as Riemannian manifold HMC (RHMC). In RHMC, the kinetic energy incorporates a second-order position-dependent metric tensor, typically the local Hessian of the potential function. Consequently, the Hamiltonian becomes non-separable in position ($x$) and momentum ($p$), rendering the generalized leapfrog updates computationally more expensive and often unstable compared to the true Hamiltonian trajectory. 

Numerical instability arises because regions of high curvature in the parameter space can induce excessively large momentum realizations, leading to unstable trajectories unless a sufficiently small step size is employed. This restriction often necessitates conservative integration settings that limit efficiency across the parameter space. Inspired by the wide use of gradient clipping in deep learning, \cite{lu:2017} proposed a modified kinetic energy formulation that effectively bounds particle velocities during Hamiltonian evolution, mitigating the risk of instability. More recently, \cite{whalley:2024} investigated the use of randomized integration times \citep{bou:2017} to enhance the numerical stability and mixing performance of RHMC. We discuss various numerical integrators and their stability properties in Section~\ref{sec:Numerical Integrators}.

\subsection{HMC for non-differentiable targets}\label{sec:non_diff_HMC}

Differentiability of the target density is a fundamental requirement for HMC. Nevertheless, Bayesian models formulated to induce structured sparsity frequently employ non-differentiable priors, thereby presenting substantial challenges in designing efficient sampling algorithms. \cite{chaari:2016} were among the first to address this issue by introducing a framework for sampling from probability distributions with non-differentiable energy functions, referred to as non-smooth HMC (ns-HMC). Their approach approximates the non-differentiable potential using a smooth envelope. Although ns-HMC provides a principled solution to handle non-smooth potentials, its computational efficiency is limited when the proximal mapping of the potential is not available in closed form and must be evaluated through iterative optimization procedures. 

More recently, \cite{shukla2025proximal} critically examined the limitations of ns-HMC and proposed the proximal HMC (p-HMC) algorithm, which integrates proximal mapping techniques within the Hamiltonian framework. They established theoretical guarantees of geometric ergodicity and offered practical guidelines for selecting tuning parameters associated with the proximal operator. Both ns-HMC and p-HMC essentially rely on smooth approximations of the non-smooth potential function via Moreau-Yosida envelopes for the purpose of gradient calculation. It remains to be seen whether there exist enveloping techniques that are particularly suited for Hamiltonian dynamics.

\subsection{HMC for multi-modal distributions}\label{sec:multi-modal_HMC}

Almost all standard MCMC methods struggle with crossing low-probability barriers that separate modes, and consequently take a long time to search for new modes or traverse from one mode to another, even in low dimensions. \cite{latuszynski:2025} mention three major challenges of multi-modal sampling, namely: a) moving between the modes, b) finding the modes, and c) sampling efficiently within the modes. Gradient-based schemes, like HMC, particularly struggle with moving between modes since local gradients tend to push the sampler towards the current local basin of attraction. 

The inability of HMC to sample effectively from multi-modal distributions has been widely acknowledged in the literature \citep{celeux:2000, sminchisescu:2007}. For example, \cite{mangoubi:2018} show that for multi-modal target distributions, the ability of HMC to transition between modes is provably worse than that of random-walk Metropolis. 

\cite{tripuraneni:2017} propose a non-canonical Hamiltonian dynamics called Magnetic HMC (MHMC), where the mechanics of the particle is coupled to a magnetic field. Magnetic dynamics accelerate the particle's transition from one mode to another. \cite{mongwe:2021} proposed perturbed
Hamiltonians designed with importance sampling. A more recent work in this context is Repelling-Attracting Hamiltonian Monte Carlo (RAHMC) by \cite{vishwanath:2024}. In RAHMC, the modified dynamics involve a friction parameter that allows (i)  mode-repelling, that encourages the sampler to move away from regions of high probability density, and (ii) mode-attracting, that facilitates the sampler's ability to find and settle near alternative modes. Moreover, \cite{vishwanath:2024} show that the RAHMC preserves various important properties of traditional Hamiltonian dynamics like volume-preservation, time-reversibility, etc.

\subsection{Numerical integrators for HMC}\label{sec:Numerical Integrators}

The main challenge in implementing the HMC method is generating the Hamiltonian trajectories themselves. Aside from a few trivial examples, we cannot solve Hamilton’s equations exactly, and any implementation must instead solve them numerically. Numerical inaccuracies, however, can quickly compromise the utility of even the most well-tuned Hamiltonian transition. An active area of research is on using alternative integrators other than leapfrog integrators \citep{leimkuhler:2004, blanes:2014} in HMC. For highly curved posteriors, stability requires a small step size $\varepsilon$, which may lead to increased computational cost. To improve accuracy without reducing $\varepsilon$, higher-order symplectic methods \citep{yoshida:1990} have been proposed, which are based on a symmetric composition of multiple leapfrog steps, achieving fourth-order accuracy. \cite{brofos:2021} study the implicit midpoint integrator in the context of RMHMC. They demonstrate that the implicit midpoint integrator exhibits superior conservation of energy, better preservation of phase-space volume and time-reversibility, compared to the standard generalized leapfrog in non-separable Hamiltonians. 

Meanwhile, \cite{pourzanjani:2019} examines HMC for multiscale distributions where the standard leapfrog step size must be extremely small. They demonstrate that an implicit integrator enables larger step sizes and improved sampling efficiency in the presence of severe geometric non-linearity in the target. On the other hand, each step requires solving fixed-point or Newton iterations, which increases the per-step cost and complexity in implementation and tuning.

\section{Discussion}
\label{sec:discussion}

The introduction of HMC has brought somewhat of a resurgence to Bayesian computation. Although not without its flaws, HMC has still been successfully employed in a wide variety of applications. Concurrently with methodological developments, there have been considerable efforts in also studying theoretical properties of HMC samplers. \cite{durmus:2017} were one of the first to study the ergodicity of the HMC algorithm. \cite{livingstone:2019} establish sufficient conditions for geometric ergodicity of HMC algorithms, with conditions similar to those required for MALA. \cite{chen2020fast} study non-asymptotic bounds on the mixing time of HMC, quantifying the improvement compared to MALA.

More recently, together with stochastic and approximate sampling schemes, HMC has found great success in computation for machine learning. The use of HMC algorithms within the machine learning community has become prevalent, particularly for sampling from high-dimensional and potentially complicated distributions. We provide a brief review of the existing machine learning literature, where HMC plays a crucial role in sampling-based inference. \cite{havasi:2018} applies Stochastic Gradient HMC (SGHMC) for inference in deep Gaussian processes, demonstrating superior posterior estimation accuracy compared to the state-of-the-art doubly stochastic variational inference framework of \cite{salimbeni:2017}. Their results illustrate the potential of Hamiltonian dynamics to provide uncertainty quantification beyond variational approximations. 

Extending this direction, \cite{foreman:2021} propose the deep-learning HMC algorithm, a generalized formulation of HMC wherein a stack of neural network layers replaces traditional leapfrog updates. This approach enables the integration of deep learning architectures into the HMC framework, allowing adaptive and learnable transition dynamics. In the context of adversarial learning, \cite{wang:2020} introduce HMC with accumulated momentum (HMC-AM), a method that adaptively controls step sizes across the trajectory. By allowing particles to evolve with variable step sizes along different directions, HMC-AM enhances flexibility in navigating complex loss surfaces common in adversarial training. 

\cite{robnik:2023} propose microcanonical HMC (MCHMC), which follows fixed energy Hamiltonian dynamics rather than the canonical ensemble used in standard HMC. Most recently, \cite{lockwood:2024} extend the classical HMC framework into the quantum domain by proposing quantum dynamical HMC (QDHMC). This hybrid approach leverages quantum computation as a proposal mechanism, replacing the classical symplectic integration step with simulations of quantum-coherent continuous-space dynamics on digital or analog quantum hardware. The QDHMC framework exemplifies an emerging intersection between quantum simulation and Bayesian computation, suggesting promising directions for scalable probabilistic inference in the quantum era.

\section{Acknowledgments}\label{sec:Acknowledgments}
The authors are thankful to Andrew Holbrook for his wonder talk available at \url{https://www.youtube.com/watch?v=Byr9JVBI9cUv} and some useful discussions that helped highlight the relationship between HMC and the MHGJ Algorithm. The authors are also thankful to Sameer Deshpande and Amy Herring for their encouragement to write this article. Finally, DV thanks Arushi Agarwal and Sarthak Parikh for creating a safe space to ask all the questions about Hamiltonian dynamics a ``physics dummy'' would ask.  

\bibliographystyle{apalike}
\bibliography{ref} 
\end{document}